\documentclass[preprint]{aastex}
\usepackage{color}

\shorttitle{EUV and Xrays on Disks}
\shortauthors{D. Hollenbach}

\newcommand{\be}{\begin{equation}}
\newcommand{\ee}{\end{equation}}

\newcommand{\ep}{\varepsilon}

\def\ltsimeq{\,\raise 0.3 ex\hbox{$ < $}\kern -0.75 em
 \lower 0.7 ex\hbox{$\sim$}\,}
\def\gtsimeq{\,\raise 0.3 ex\hbox{$ > $}\kern -0.75 em
 \lower 0.7 ex\hbox{$\sim$}\,}

\let\gta=\gtsimeq
\let\lta=\ltsimeq
\begin{document}

\title{DIAGNOSTIC LINE EMISSION FROM EUV AND X-RAY ILLUMINATED DISKS
AND SHOCKS  AROUND LOW MASS STARS}

\author{David Hollenbach$^1$, U. Gorti$^{1,2}$}

\affil{$^1$SETI Institute, 515 North Whisman Road, Mountain View, CA 94043}

\affil{$^2$NASA Ames Research Center, Moffett Field, CA 94035}

\begin{abstract}

Extreme ultraviolet (EUV, 13.6 eV $< h\nu \lta 100$ eV) and X-rays in the 0.1-2 keV band can 
heat the surfaces of disks around young, low mass stars to thousands of degrees and ionize species with
ionization potentials greater than 13.6 eV.   Shocks generated by protostellar winds can also
heat and ionize the same species close to the star/disk system.  These processes produce diagnostic lines
(e.g., [NeII] 12.8 $\mu$m and [OI] 6300 \AA )
that we model as functions of key parameters such as  EUV luminosity and spectral shape, 
X-ray luminosity and spectral shape, and wind mass loss rate and shock speed.
Comparing our models with observations,
we conclude that either internal shocks in the winds or X-rays incident on the disk surfaces often produce the observed [NeII]  line, although there are cases where EUV may dominate.    Shocks
created by the oblique interaction of  winds with 
disks are unlikely [NeII] sources  because these shocks are too weak to ionize Ne.
Even if  [NeII] 
is mainly produced by X-rays or internal wind shocks, the neon observations typically  place upper limits of  $\lta 10^{42}$ s$^{-1}$  on the EUV photon
luminosity of these young low mass stars.
The observed [OI] 6300 \AA \ line has both a low velocity component (LVC)  and a high velocity component.  The
latter likely arises in internal wind shocks.   For the former we find that X-rays likely produce more [OI] luminosity than either the EUV layer, the transition layer between the EUV and X-ray layer, or the shear layer where
the protostellar wind shocks and entrains disk material in a radial flow across the surface of the disk.  Our soft X-ray models
produce [OI] LVCs with luminosities up to $10^{-4}$ L$_\odot$, but may not
be able to explain the most luminous LVCs.

\end{abstract}

\section{INTRODUCTION}

The photoevaporation of a protoplanetary disk by the extreme ultraviolet
(EUV, 13.6 eV $< h\nu \lta 100$ eV) or Lyman continuum photons from  the
central star may significantly affect the formation and evolution of planets
and planetesimals, and may be one of the important mechanisms for dispersing
disks (Hollenbach et al 1994,  Hollenbach, Yorke \& Johnstone 2000, Clarke et al 2001, 
 Richling, Hollenbach, \& Yorke 2006, Alexander et al. 2006a,b, Alexander 2008a).  
 EUV photoevaporation occurs because
 the EUV photons create a $10^4$ K, ionized surface on the disk, and beyond
about 1($M_*/1\ $M$_\odot$) AU, where $M_*$ is the stellar  mass,  the thermal pressure of the gas is
sufficient to drive a significant hydrodynamic flow out of the gravitational potential
of the star and into interstellar space.

Some of the most detailed models
of the dispersal of disks around isolated low mass stars invoke
viscous spreading and accretion on the inside ($\lta$ few AU) of the disk
and EUV-induced photoevaporation on the outside (Clarke et al 2001,
Matsuyama et al 2003, Ruden 2004, Alexander et al 2006 a,b). This combination
has been invoked to explain gas-poor giants like Uranus and Neptune (Shu,
Johnstone, \& Hollenbach 1993), the rapid evolution of classical T Tauri
stars to weak-lined T Tauri stars (Clarke et al 2001, Alexander et al
2006 a,b), the production of large inner holes such as exist in
some sources (Alexander et al 2008b, Cieza et al 2008), and the migration 
and ``parking'' of giant planets (Matsuyama et al 2003, Lecar \& Sasselov 2003,
 Veras \& Armitage 2004).

X-rays from the star also significantly affect the disk.  Glassgold et al
(2004) show that hard X-rays can penetrate to moderate depths
into the disk and produce sufficient ionization to maintain a vigorous
magnetorotational instability (MRI, Balbus \& Hawley 1991), at least in
the upper layers of the disk (see also, Sano et al 2000, Stone \& Pringle 2001).  
Chiang \& Murray-Clay (2007) have recently
expanded on this idea, using X-rays to stimulate MRI in the inner edge
of a dusty disk, thereby eating away the disk from inside out.
Alexander et al (2004a) have argued that X-rays by themselves do
not lead to significant photoevaporation, but Gorti \& Hollenbach (2009)
have shown that $\sim 1$ keV X-rays may increase FUV-induced photoevaporation rates
by roughly a factor of 2, because X-ray ionization increases the electron
abundance, which enhances the FUV grain photoelectric heating mechanism.
More recently (Ercolano et al 2008, 2009, Gorti, Dullemond, \& Hollenbach 2009),
it has become clear that a soft (0.1-0.5 keV) X-ray component can lead to
significant photoevaporation rates.
Glassgold, Najita, \& Igea (2007) and Meijerink et al (2008) have shown that X-rays
partially ionize and heat the gas just below the EUV fully ionized layer,
and that the X-ray heated gas achieves temperatures of order 1000 to
4000 K in a dense ($n \sim 10^7$ cm$^{-3}$) layer out to about 10-20 AU.  Although they do not
discuss X-ray photoevaporation, these temperatures, densities and radii suggest
significant rates.

EUV and X-ray photons around low mass stars, whose photospheres are too
cool to produce a substantial EUV or X-ray flux, emanate from the accretion shock
created
by the impact of the accreting disk gas onto the stellar surface and/or from
the hot plasma generated by
magnetic activity on the stellar surface akin to (but much greater than)
the Sun's chromosphere or corona. These two mechanisms heat plasma to
temperatures $>> 10^4$ K,
 and thereby produce significant EUV and X-ray luminosity.
Alexander et al (2004b) argue that the EUV photons do not
likely penetrate the accretion columns to irradiate the disk, and that,
therefore, magnetic activity is a more attractive source for the EUV
that shines on the disk surface. However, many accreting sources exhibit a soft X-ray component
(e.g., Kastner et al. 2002, Stelzer \& Schmitt 2004), which may arise from
an accretion shock or is at least mediated by accretion flows (Preibisch 2007,
G\"{u}del \& Telleschi 2007).
Soft X-rays are only
somewhat more penetrating than EUV photons, raising the possibility
that the geometry of the accretion streams (sometimes called "funnel flows")
onto the star may also allow the escape of at least some of these hydrogen-ionizing photons.
 The hard ($\gta 1$ keV) X-rays
likely arise from the magnetic activity (i.e., the chromosphere and
corona).

Whatever the source of EUV photons, they must still penetrate
protostellar winds.  Protostellar winds are thought to be driven by 
magnetohydrodynamic processes from the inner portions of accreting disks 
(e.g., Shu et al. 1994, Ouyed \& Pudritz 1997). 
We show in \S 2 that these winds must have low mass loss rates,
$\dot M_w \lta 10^{-9}$ M$_\odot$ yr$^{-1}$, for EUV or soft ($\lta 0.2$ keV) X-rays
 to penetrate
them and to illuminate the disk surface beyond $\sim 1$ AU, where photoevaporation
proceeds.   Since accretion rates onto the central star are correlated with
protostellar wind mass loss rates (Hartigan et al. 1995, White \&
Hillenbrand 2004), this critical wind mass loss rate corresponds
to an accretion rate of about $10^{-8}$ M$_\odot$ yr$^{-1}$.

The main weakness in EUV photoevaporation models is the extreme
uncertainty in the EUV photon luminosity $\Phi _{EUV}$ of the central
star. The EUV opacity of  hydrogen is so high that
a column of only $\sim 10^{17}$ hydrogen atoms/cm$^2$ provides optical depth
of order unity.
 Therefore, interstellar extinction prevents the direct
observation of the EUV flux from young, low mass stars with disks.
There are, however, observations of nearby, older, solar-mass stars,
including the Sun, which provide a clue to the spectra from the far
ultraviolet (FUV, 6eV $< h\nu < 13.6$ eV) to the X-ray of low mass stars
due to their magnetic activity (Ribas et al 2005). These suggest that,
very roughly, $\nu F_{\nu}$ is constant for a given star from the FUV band
through the EUV band to the keV X-ray band.  Ercolano et al (2009) also discuss
observations of flare stars which suggest magnetically-heated coronae on the stellar surfaces 
with a range of plasma temperatures resulting in roughly  $F_\nu \propto \nu ^{-1}$ power law EUV spectrum.
Thus, one might estimate the magnetically-produced EUV luminosity of a low mass star
by measuring either (or both) the 0.1-1.0 keV X ray luminosity or
the 6 - 13.6 eV FUV luminosity.  For non-accreting but young ($\sim$
1 Myr) low mass stars, the X ray and FUV luminosities tend to be of order
$\sim 10^{-3}$ L$_{bol}$ (e.g., Flaccomio et al. 2003, Valenti et al. 2003),
 suggesting $L_{EUV} \sim 10^{-3}$ 
 L$_\odot$ or $\Phi _{EUV} \sim 10^{41}$ EUV photons per second
for a 1 M$_\odot$ star. 

Alexander et al (2005), based on earlier work of Brooks et al (2001),
 use FUV emission lines of various ions of the elements
carbon, oxygen, nitrogen and silicon seen in T Tauri stars
to try to estimate the distribution of emission measures as a function
of T of the hot ($\sim 10^4 - 10^6$ K)
plasma.   As the authors themselves
point out, this method is fraught with difficulties, and, as a result,
they can only constrain $\Phi _{EUV}$ to range from $10^{41} - 10^{44}$ s$^{-1}$
in young solar mass stars.   The knowledge of the EUV luminosity is
critical in predicting EUV-driven photoevaporation and determining whether it
dominates disk evolution and explains the observed short ($\sim 1-3$ Myr)
lifetimes of disks around low mass stars.

One way to measure $\Phi _{EUV}$ is to observe emission lines produced by
the heating and ionization caused by 
these photons on the disk surface.  Such a measurement is important since
$\Phi _{EUV}$ determines the EUV photoevaporation rates, and therefore
the EUV-induced dispersal times of the gas and dust in these young, planet-forming
disks.  Given a disk illuminated by EUV photons, a tenuous, $10^4$ K, fully
ionized surface is created by the photoionization of hydrogen.  In effect,
a sort of ``blister HII region'' is created above the bulk of the disk, which is
mostly neutral molecular gas.  Although this HII region contains very little 
mass ($\sim 10^{-7}[\Phi_{EUV}/10^{41}$ s$^{-1}$]$^{1/2}$M$_J$, where M$_J$ is a Jupiter
mass), it can produce sufficient forbidden optical line emission (e.g., [SII]
6731 \AA \ and [NII] 6583 \AA , see Font et al
2004) or infrared fine structure emission (e.g., [NeII] 12.8 $\mu$m, this paper) to be observed.
We note that [NeII] 12.8 $\mu$m is one of the strongest lines from HII regions associated
with Giant Molecular Clouds, and, because neon is not depleted and its gas phase
abundance relative to hydrogen is quite well known, this fine structure line can
also be used in these regions to measure or constrain the ionizing luminosity of the exciting star(s)
(Ho \& Keto 2007).

There are two problems in using the emission lines from the HII surface to measure
$\Phi _{EUV}$.  Uncertainties in extinction, the gas temperature and the gas density
make the {\it optical} lines a poor diagnostic of $\Phi _{EUV}$.  The infrared fine
structure lines are much better for this purpose, but they also can be produced
by the heating and (partial) ionization of the neutral gas below the HII surface
by penetrating X rays (Glassgold, Najita, \& Igea 2007).  In
addition, they can be produced in high velocity (ionizing) shocks created by
the protostellar wind. 
We discuss in this paper the relative contributions to the fine structure emission by
the surface EUV heated layer, the subsurface X ray heated layer, and the wind shocks.
However, even if its origin cannot be distinguished, the fine structure emission,
e.g., [NeII] 12.8 $\mu$m, gives a strict upper limit on $\Phi _{EUV}$.   In addition,
if arising from the EUV or X-ray layers, 
the [NeII] and other fine structure lines provide a measure of the density and temperature 
of the hot surface gas, and therefore directly probe some of the regions where
photoevaporation originates (Alexander 2008).

 [NeII] 12.8 $\mu$m emission from young stars with optically thick disks was first detected using the high resolution mode of the IRS instrument on the Spitzer Space Telescope (Pascucci et al. 2007, Lahuis et al. 2007, Ratzka et al. 2007, Espaillat et al. 2007), and is now found in over $\sim 50$ sources (G\"udel et al 2009). Some of these sources ($\sim 15$) also show emission from  the hydrogen recombination lines H(7-6)$\alpha$ and H(6-5)$\alpha$, and only one source is detected in [NeIII]15$\mu$m. Observed line luminosities range from  $10^{-4}-10^{-6}$ L$_{\odot}$.  Follow-up, very high resolution ground-based observations of some bright [NeII] sources have resolved the line emission and observed line widths ($\sim15-80$km s$^{-1}$), interpreted as emission arising from X-ray heated layers in  Keplerian-rotating disks (Herczeg et al. 2007, Najita et al. 2009), EUV photoevaporative flows (Herczeg et al. 2007, Pascucci \& Sterzik 2009) or from outflows associated with these sources (van Boekel et al. 2009, Najita et al. 2009). Correlations have been sought  between the [NeII] luminosities and disk and stellar diagnostics such as X-ray luminosity (Pascucci et al. 2007,
 G\"udel et al 2009) and mass accretion rates (Espaillat et al. 2007, G\"udel et al 2009), but the data is inconclusive. The origin of the [NeII] emission, although widely attributed to disks, is still not definitive. 

This paper is motivated by the recent observations of [NeII] 12.8$\mu$m emission.   We model disks illuminated by EUV and X rays, and present results for the infrared fine structure lines
of  Ar$^+$, Ar$^{++}$,  Ne$^+$,
Ne$^{++}$, N$^+$, N$^{++}$, O$^{++}$, S$^{++}$, and S$^{+++}$,
 two infrared recombination lines
of H$^+$, and the optical forbidden line [OI] 6300 \AA.   We show that if the EUV layer dominates the emission, the infrared fine stucture lines 
diagnose $\Phi _{EUV}$ and the shape (slope) of the EUV spectrum.
We also show that measurements of [NeII] 12.8 $\mu$m and [NeIII] 15$\mu$m 
 are particularly good diagnostics of these parameters, being strong and relatively 
 insensitive to extinction and changes
in the plasma density $n$ or temperature $T$. Our models of the X-ray layers, like
 the X-ray models of Glassgold et al (2007) and Meijerink et al (2008),
produce [NeII] emission that, at least in some cases, is in accord
with the observations. However, in a number of cases the X-ray
heating mechanism seems insufficient to provide the emission (Espaillat et al. 2007, G\"udel
et al 2009),
as we will also show in this paper.  Shocks in the protostellar wind or an unseen
EUV or soft ($\sim 0.1-0.3$ keV) X-ray component may provide the origin
of the [NeII] in these cases.
Our models differ from Glassgold et al 
in that we treat the vertical structure
of the disk consistently (that is, the gas temperature is not assumed to
equal the dust temperature in calculating the vertical density structure), 
 include EUV ionization and heating, include FUV photodissociation and heating,
 treat the X-ray heating somewhat 
differently, and include some additional
significant cooling lines, such as [NeII] 12.8 $\mu$m and [ArII] 7 $\mu$m.  

This paper complements earlier (Gorti \& Hollenbach 2004, 2008)
papers which examined the molecular and atomic
fine structure emission from the neutral disk.  In this older work, 
the fine structure lines treated  focussed mainly on those with ionization potentials
less than 13.6 eV, such as those of O, C, C$^+$, S, Si, Si$^+$, Fe, and Fe$^+$, although
we did treat the X-ray ionization of some species in the predominantly neutral gas.
In this paper we focus on species with ionization potentials greater than
13.6 eV, which are only found in the fully photoionized HII region surfaces of disks,
 in X-ray ionized, predominantly neutral gas, or in fast ($\gta 100$ km s$^{-1}$)
 shocks produced by the stellar wind.

 We organize the paper
as follows.  We discuss the restriction on the wind mass loss rate
in order for  the FUV, EUV and X-ray radiation from the protostar to penetrate the wind and
shine on the disk surface in \S 2.  Section 3   provides
analytic estimates of the relation of the fine structure 
and hydrogen recombination line luminosities
to $\Phi _{EUV}$, the scaling of the emission from the X-ray layer to the 
X-ray luminosity of the central star, the [NeII] luminosity produced in
wind shocks, and the [OI] 6300 \AA \ luminosity possible from both the disk
and from wind shocks.  Section 4  shows the results of numerical models. 
Section 5 compares the results of our models to recent observations
made by the Spitzer Space Telescope and several ground-based telescopes, and discusses
the relative contributions of EUV, X rays  and shocks to the observed [NeII], hydrogen
recombination lines, and [OI] emission.
  We conclude with a discussion and summary  in \S 6.

\section{FUV, EUV AND X-RAY PENETRATION OF PROTOSTELLAR WINDS}

Although our protostellar wind model is influenced by the ``X wind'' models
of Shu et al (1994), the main assumption we make is that the bulk of the
wind mass loss rate $\dot M_w$
originates from cylindrical radius $r_w$ to $r_w + fr_w$, where $r_w \sim 
10^{12}$ cm and $f \sim 1$. Therefore, the model applies also to other disk wind 
models (e.g., Ouyed \& Pudritz 1997) where the bulk of the mass loss originates
from the inner disk surface. We assume that $f$ is sufficiently small that we can
take $n_b$ as the average hydrogen nucleus density at the base of the wind without
introducing significant error by assuming this constant density from $r_w$
to $r_w + fr_w$.

The mass loss in the wind, $\dot M_w$, arising from this geometry is given
\begin{equation}
\dot M_w = 2\left[\pi (r_w +fr_w)^2 - \pi r_w^2\right]n_bm_Hv_w,
\end{equation}
where $m_H=2.3 \times 10^{-24}$ gm is the mass per hydrogen nucleus and $v_w$ is
the wind velocity.
The hydrogen nucleus column density $N_w$ through the base of the wind, which
the energetic photons must penetrate to reach the outer disk surface, is then
given:
\begin{equation}
N_w \simeq n_bfr_w \simeq 2.2 \times 10^{21}\left({{\dot M_w}\over{10^{-8}\ {\rm M}
_\odot \ {\rm yr}^{-1}}}\right)\left({{100\ {\rm km\ s}^{-1}}\over {v_w}}\right)
\left({{10^{12} \ {\rm cm}}\over {r_w}}\right)\left({1 \over {1+0.5f}}\right) \ 
{\rm cm}^{-2}.
\end{equation}
Interstellar dust requires a hydrogen nucleus
column of $\sim 10^{21}$ cm$^{-2}$ to provide optical depth unity in the FUV.
However, the dust lifted from the surface of the disk at the base of the wind
is likely to have coagulated to much larger sizes than interstellar dust, and
furthermore to have lower dust/gas mass ratios because of sublimation of the 
less refractory materials and settling of the refractory grains to the
midplane (Dullemond \& Dominik 2005).   In fact, at radii of $\lta 10^{12}$  cm
it is possible that all dust has sublimated.   All these processes lower the
dust cross section per hydrogen nucleus.   Even if there is no dust (for example,
if all the dust is sublimated), the gas
provides FUV opacity and attenuates the FUV significantly for columns greater
than about $10^{24}$ cm$^{-2}$.  Assuming a minimum reduction in dust opacity
relative to interstellar dust of a factor of 10, FUV
will penetrate wind columns $N_w \lta 10^{22}$ cm$^{-2}$.  Dust also provides a
source of X-ray opacity, which will be reduced from interstellar values by
the effects of settling and coagulation.  However, considerable opacity
remains in the gas phase elements such as C, O and Ne.  Gorti \& Hollenbach
(2004, 2008) estimate, using the cross-sections of Wilms, Allen \& McCray(2000),
 that $N_w \sim 10^{22}$ cm$^{-2}$  is required for 1 keV optical 
depth unity at disk surfaces.  On the other hand, soft X-rays experience considerably
more optical depth, and $N_w \sim 10^{20}$ cm$^{-2}$ provides optical depth
unity for $\sim 0.2$ keV X-rays.  Therefore, in summary, 
$\dot M_w \lta 4 \times 10^{-8}$ M$_\odot$ yr$^{-1}$ is required for $\sim 1$ keV X rays 
to penetrate the protostellar wind, whereas soft X-rays can only penetrate when
$\dot M_w \lta 4 \times 10^{-10}$ M$_\odot$ yr$^{-1}$.  The penetration of the FUV
likely occurs at mass loss rates considerably higher than $\dot M_w \sim
4 \times 10^{-8}$ M$_\odot$ yr$^{-1}$ because of dust sublimation and settling,
but this number serves as a useful lower limit.

A column $N_w$ of 10$^{20}-10^{22}$ cm$^{-2}$ of {\it neutral} hydrogen is totally
opaque to EUV photons, since $N(HI) \sim 10^{17}-10^{18}$ cm$^{-2}$ produces 
EUV optical depth unity. For EUV photons to penetrate the wind, the EUV photon
flux $F_{EUV}$ must 
be sufficiently high to keep the base of the wind fully ionized, so that
$n(HI)/n_b <<1$ and $N(HI) \lta 10^{17}$ cm$^{-2}$.  This ``Str\"omgren''
condition
\begin{equation}
F_{EUV} > \alpha _{r,H}n_b^2 fr_w,
\end{equation}
can be rewritten as
\begin{equation}
\dot M_w \lta 8 \times 10^{-10} \left({{fr_w}\over {10^{12} \ {\rm cm}}}\right)^{1/2}
\left({{\Phi _{EUV}}\over {10^{41} \ {\rm s}^{-1}}}\right)^{1/2}\left({v_w \over
{100 \ {\rm km\ s}^{-1}}}\right) \ \ {\rm M_\odot \ yr^{-1}}.
\end{equation}
In other words, the mass loss rate has to be less than the RHS of this equation
for EUV to penetrate the base of the wind and illuminate the outer disk surface
beyond 1 AU.

Wind mass loss rates are hard to measure ``directly'' from the observed optical
line emission (e.g., [SII]) seen in their jets.  The derived mass loss rates
from these optical lines depend on knowing the gas temperature, the gas density,
and the ionized fraction -- all of which are quite uncertain.  ``Indirect'' methods
rely on measuring the momentum in swept up circumstellar gas.  This method
is also approximate since it requires an estimate of the wind speed, the
duration of the mass loss episode,  and the
conversion factor of CO luminosity to mass.  Likewise there are uncertainties
in observationally determining the mass accretion rate $\dot M _{acc}$
onto the central star.  These uncertainties create a spread in the constant $k$
of proportionality, but it is generally agreed that the wind mass loss rate
scales with the mass accretion rate, $\dot M_w \simeq k \dot M_{acc}$.  The
constant $k$ has been estimated from $\sim 0.01$ (Hartigan et al 1995) to $\sim 0.1$ (White
\& Hillenbrand 2004).
The Shu et al (1994)  X wind model predicts values somewhat higher than 0.1.
White and Hillenbrand point out that there seems to be considerable {\it intrinsic}
scatter in the ratio of wind mass loss rate to mass accretion rate from source to source.

Roughly then, if we take $\dot M_w \sim 0.1 \dot M_{acc}$, the FUV and $\sim 1$
keV X-rays 
penetrate the wind when $\dot M_{acc} \lta 4 \times 10^{-7}$ M$_\odot$ yr$^{-1}$,
whereas the EUV and soft ($\sim 0.2$ keV) X-rays penetrate the wind when the accretion rate has dropped to
$\dot M_{acc} \lta 8 \times 10^{-9}$ M$_\odot$ yr$^{-1}$.  Hartmann et al (1998)
show the evolution of $\dot M_{acc}$ for young, solar mass stars.  With order
of magnitude dispersion, $\dot M_{acc}$ is roughly $10^{-8}$ M$_\odot$ yr$^{-1}$
at 1 Myr, and drops rapidly on Myr timescales. Thus, FUV and $\sim$ 1 keV X rays
illuminate the disk surface nearly as soon as the epoch of heavy accretion
of material onto disk and star from the natal cloud core has ceased.  However,
EUV and soft X-rays may not illuminate the disk surface until roughly 1-2 Myr 
has elapsed from that time.

If one wishes to observe a disk whose ionized fine structure lines are not produced
 by  EUV and soft X-rays, one should select sources with $ \dot M_{acc} \gta 8 \times 10^{-9}$ 
M$_\odot$ yr$^{-1}$. If [NeII], for example, is
detected in sources with    $8 \times 10^{-9}$ 
M$_\odot$ yr$^{-1} \lta \dot M_{acc}\lta  4 \times 10^{-7}$ M$_\odot$ yr$^{-1}$, then
hard ($\sim 1$ keV) X-rays or possibly wind shocks may be implicated.  If [NeII] is detected in sources
with  $\dot M_{acc}\gta  4 \times 10^{-7}$ M$_\odot$ yr$^{-1}$, then protostellar wind shocks
almost certainly provide the origin. 
If one wishes to observe sources illuminated by both X-rays and EUV, and therefore
containing, for example, [NeII] emission from an EUV-produced HII surface and
also from an X-ray-produced partially ionized deeper layer, then one should
observe sources with $\dot M_{acc} \lta 8 \times 10^{-9}$ M$_\odot$ yr$^{-1}$.
Interestingly, the [NeII] sources have been detected in [NeII] emission in this entire range
of   $\dot M_{acc}$, suggesting a wide range of origin of the [NeII] [see Espaillat
et al (2007) and  G\"udel et al (2009)].

\section{ANALYTIC MODELS OF EMISSION LINES DIAGNOSTIC OF EUV AND X-RAYS INCIDENT ON DISKS OR OF SHOCKS}

There are basically three types of lines diagnostic of EUV and X-rays incident on disks or of fast, ionizing shocks
in protostellar winds: hydrogen
recombination lines, optical forbidden
lines like [OI] 6300 \AA , [SII] 6713 and  6731 \AA,  and [NII] 6583 \AA ,  and the infrared fine structure lines
of ionized species whose ionization requires photons more energetic than 13.6 eV.   We discuss here
the last two, and leave the discussion of hydrogen recombination lines for \S 3.2.   With the possible
exception of [OI] 6300 \AA, the optical lines likely arise in the completely ionized HII region at
the surface of the disk or in fast, ionizing shocks because the lines typically lie $\gta 20,000$ K above ground and
are excited mainly by electron collisions; these regions have higher temperatures 
($\sim 10^4$ K versus several thousand K in the X-ray heated region) and generally higher
electron densities than the X-ray layers.   In addition, most of the optical lines arise from ionized species whose abundances
peak in the completely ionized HII gas as opposed to the mostly neutral X-ray layers (the notable exception being [OI]).  
 The infrared fine structure lines from high IP ($>13.6$ eV) species 
typically lie $\sim 300-1000$ K above
ground and therefore are not sensitive to temperature for temperatures above about $300 - 1000$ K.
These lines may come from either the EUV-heated HII region, the X-ray heated region, or shocked regions
and we show below that  the relative EUV versus X-ray photon luminosity
from the central star determines which of these two regions will dominate the emission.    We focus in this section on
the infrared lines, because Font et al (2004) and Meijerink et al (2008) have discussed
the optical emission from the EUV and X-ray heated layers.  However, we do include
the [OI] 6300 \AA \  line in our analysis, because our [OI] luminosities from the X-ray layer differ 
from the Meijerink et al values, and because other researchers have not been able
to match the observed luminosities in this line (Hartigan et al 1995).  We also include a discussion of the infrared
hydrogen recombination lines that have been observed from these star/disk systems.

\subsection{Fine Structure Lines from the HII Surface (EUV Layer)}
 Consider an axisymmetric disk 
described by cylindrical coordinates $r$, $z$. If $f_{EUV}$ is the fraction
of ionized photons from the star absorbed by the disk, then the Str\"omgren
condition is:
\begin{equation}
f_{EUV}\Phi _{EUV} = 2 \alpha _{r,H} \int _{z_{IF}}^{\infty} dz \int _{r_i}^{r_o} 2 \pi
r n_e^2 dr,
\end{equation}
where the electron density $n_e$ is a function of $r$ and $z$ but is negligible
below $z_{IF}$, the ionization front, and where we have ignored dust attenuation
in the HII surface region above $z_{IF}$.  We will justify the neglect of dust
post facto below, as well as show that $f_{EUV} \sim 0.7$.
 In  Eq. (5)
$r_i$ and $r_o$ are the inner and outer 
radii of the disk, and $\alpha _{r,H}$ is the case B recombination coefficient
for electrons with protons ($\alpha _{r,H} = 2.53\ \times 10^{-13}$ cm$^3$ s$^{-1}$
at $T= 10^4$ K, Storey \& Hummer 1995). 

For simplicity we treat the specific example of a simple two level fine structure
system like [NeII] 12.8 $\mu$m.  Then,
for a transition $t$  we have that the escaping line
luminosity $L_t$ from the disk is given:
\begin{equation}
L_t \simeq  \gamma _t \Delta E_t \int _{z_{IF}}^{\infty} dz \int _{r_i}^{r_o}
2 \pi r \left({{n_e n(i)} \over {1+{n_e \over n_{ecr,t}}}}\right) dr,
\end{equation}
where $\gamma _t$ is the collisional excitation rate coefficient of the transition, $\Delta E_t$
is the photon energy, $n_{ecr,t}$ is the critical electron density for the transition,
and $n(i)$ is the density of the ionized species $i$ that produces the transition.  Note that here
we have included the fact that half the emitted infrared photons are directed
toward the disk midplane, where they are absorbed by the (assumed) optically thick disk,
and half escape.  If we set $n(i)=x_s f(i)n_e$, where $x_s$ is the  
abundance of species $s$ (in all ionization states) in the EUV layer, and $f(i)$ is the fraction of that species in ionization
state $i$, then  we can write  using Eq.(5):
\begin{equation}
L_t= \gamma _t  \Delta E_t \left( {{f_{EUV} \Phi _{EUV} x_s f(i)}\over {2 \alpha _{r,H}}}\right)
\left({1 \over {1+ {n_e \over n_{ecr,t}}}}\right).
\end{equation}
We note that in taking $x_sf(i)$ out of the integral in Eq.(6) we implicitly assume that in Eq.(7)
$x_sf(i)$ is the density weighted average of this product in the EUV layer.
Thus, we find that if $n_e < n_{ecr,t}$ and if $f(i)$ does not depend on $\Phi _{EUV}$ (e.g, for the
dominant ionization state where $f(i)\simeq 1$), the line luminosity $L_t$ is directly
proportional to the ionizing photon luminosity $\Phi _{EUV}$ of the central star.
If EUV is the sole excitation source, the measurement of $L_t$ directly 
measures the uncertain parameter $\Phi _{EUV}$ if we have a knowledge of
$x_s$ and $f(i)$.   The main unknown is the fraction $f(i)$ in a particular 
ionization state $i$, since the gas phase abundance of an element can often be estimated from 
observations of HII regions.  For example, neon often is found as Ne$^+$ and Ne$^{++}$.
We therefore rewrite Eq. (7) as
\begin{equation}
L_t= C_t f(i) f_{EUV} \Phi _{EUV}
\left({1 \over {1+ {n_e \over n_{ecr,t}}}}\right),
\end{equation}
where $C_t \equiv \gamma_t  \Delta E_t x_s/(2 \alpha _{r,H})$.  Note that $C_t$
and $n_{ecr,t}$ are known quantities and therefore constants in the equation.
 If the transition is from the dominant
ionization state of the species, then $f(i) \sim 1$.  Our modeling of disks suggests
$f_{EUV} \sim 0.7$.  $C_t$ is therefore the main constant of proportionality
that links the observed IR line luminosity to the EUV photon luminosity; for
$n_e < n_{ecr,t}$ and taking the dominant ionization state, $C_t$ is the 
energy per absorbed EUV photon that emerges from the disk in the fine structure transition.
Table 1 lists the value of $C_t$ for the various transitions
considered in this paper, along with $n_{ecr,t}$ and the assumed $x_s$ that is
used in $C_t$. 

The above equations show that it is important to estimate the electron density in
the ionized surface of the disk at the radius where most of the emission in a given
line is produced. We follow
the results of Hollenbach et al (1994). Because the HII region surface is
isothermal, the electron density at a given $r$ decreases from $z_{IF}$ upwards
as 
\begin{equation}
n_e(r,z)=n_e(r, z_{IF}) e^{-\left({{z-z_{IF}}\over {2H}}\right)^2},
\end{equation}
where $H$ is the isothermal scale height of the $10^4$ K gas given by
\begin{equation}
H(r) = r_g\left( {r \over r_g}\right)^{3/2}
\end{equation}
for $r < r_g$, and where $r_g$ is given by
\begin{equation}
r_g \simeq 7 \left({M_* \over {1\ {\rm M}_\odot}}\right) \ {\rm AU}.
\end{equation}
Note that in Eq.(9) both $z_{IF}$ and $H$ are functions of $r$.
The gravitational radius $r_g$ is where the sound or hydrogen thermal speed
is equal to the escape speed from the gravitational potential of the star.
For $r > r_g$, where the gas is freely evaporating,
the effective height of the disk is $H \sim r$, since the initially vertically
flowing gas turns over on radial streamlines by the time 
$z=r$.
Hollenbach et al also showed that the base electron density at $z_{IF}$ falls
off radially as a power law.
\begin{equation}
n_e(r, z_{IF}) \simeq 5 \times 10^4 \left({{\Phi _{EUV}}\over {10^{41} 
\ {\rm s}^{-1}}}\right)
^{1/2}\left({r_g \over r}\right)^{3/2} \left( {{ 1 \ {\rm M}_\odot}\over
{M_*}}\right)^{3/2} \ {\rm cm}^{-3} \ \ {\rm for} \ r < r_g
\end{equation} 
and
\begin{equation}
n_e(r, z_{IF}) \simeq n_e(r_g, z_{IF}) \left( {r_g\over r}\right)^{5/2} \ \ {\rm for} 
\ r > r_g.
\end{equation} 
Note that if $n_e(r, z_{IF}) < n_{ecr}$ everywhere, then the amount of luminosity
$L(r)$ from a logarithmic interval of $r$ is proportional 
to the volume emission measure $n_e^2 V$ at that 
$r$, or to $n_e^2 r^2 H$.  For $r<r_g$ we then obtain $L(r) \propto r^{1/2}$
whereas for $r>r_g$ we obtain $L(r) \propto r^{-2}$.  In other words, the line
luminosity originates mostly from $r \sim r_g$.  If the electron densities in the
very inner regions exceed $n_{ecr}$ but are less
than $n_{ecr}$ at $r_g$, the luminosity is relatively unaffected, and our conclusion
does not change.  However, if  $n_e(r_g, z_{IF}) > n_{ecr}$, then the line
luminosity will drop.  The line emissivity will be suppressed by a factor of approximately
 $n_e(r_g, z_{IF})/n_{ecr}$ at $r_g$.  However, now $L(r) \propto n_eV$ as long
as $n_e > n_{ecr}$, so that $L(r) \propto r^{1/2}$ beyond $r_g$ until the
density drops below the critical density, when it reverts to its former $r^{-2}$
dependence.  The luminosity of a low critical density species will therefore
originate from the radius 
\begin{equation}
r_{max}= \left({{n_e(r_g, z_{IF})}\over {n_{ecr}}}\right)^{2/5} r_g,
\end{equation}
where $n_e(r_{max}, z_{IF})= n_{ecr}$.  As a result,
\begin{equation} L \propto n_e(r_{max},z_{IF}) r_{max}^3 \propto
\Phi _{EUV}^{3/5}
\end{equation}
 as long as $n_e(r_g,z_{IF}) > n_{ecr}$ and $f(i)$ is constant.

As examples, consider [NeII] 12.8 $\mu$m and [SIII] 19 $\mu$m.  
The critical density for [NeII] is given
$n_{ecr,[NeII]}\equiv A_{21}/\gamma_{21} \simeq 6 \times 10^5$ cm$^{-3}$, where we have taken
$A_{21}= 0.00859$ s$^{-1}$ and a collisional de-excitation rate coefficient
at $10^4$ K of $\gamma _{21} = 1.355 \times 10^{-8}$ cm$^3$ s$^{-1}$ (Griffin et al. 2001).
Thus, comparing this critical density with the electron density at $r_g$ [Eq. (12)],
we see that [NeII] is subthermal at $r_g$ typically, and that therefore it 
mostly originates from $r_g$ and tracks $\Phi _{EUV}$ linearly as long as $f(Ne^+)$ is
constant.  However,
[SIII] has a critical density of approximately $5 \times 10^3$ cm$^{-3}$ which
is about 10 times less than  $n_e(r_g, z_{IF})$ for a solar mass star with
$\Phi _{EUV} \sim 10^{41}$ s$^{-1}$(see
Eq. 12). Therefore, $r_{max} \simeq 10^{2/5} r_g
\simeq 2.5 r_g$. The [SIII] line luminosity is down
by a factor $10^{4/5} \sim 6.3$ from the value it would have had if it had
been subthermal at $r_g$, rather than the factor of 10 drop at $r_g$,
because most of the emission comes from further out where there is more
volume.  The luminosity in the line will not linearly track $\Phi _{EUV}$
 because of the significant (and variable with $\Phi _{EUV}$) collisional
de-excitation of the transition.   In fact, as long as $n_e(r_g,z_{IF}) > n_{ecr}$
{\it and $f(S^{++})$ is not dependent on $\Phi_{EUV}$},
the luminosity in the line scales as $\Phi_{EUV}^{3/5}$ as shown in Eq. (15).  However, we find in
our numerical analysis that $f(S^{++})$ does depend significantly on $\Phi_{EUV}$, 
and the  $\Phi_{EUV}^{3/5}$ dependence is not seen (as we show below in our numerical
models).

We have shown above that the fine structure emission from the HII surface
region arises from $\sim r_g$ as long as $n_e(r_g,z_{IF}) < n_{ecr}$.   However, this
conclusion and the important analytic derivation of the line luminosities [Eq. (8)] both
require that the surface HII region be ionization-bounded (i.e., that there be a neutral layer
underneath the completely ionized surface).
The minimum mass of gas required to
fully absorb the incident EUV photons and create an ionized surface region
with a neutral midplane region for $r<r_g$ is given:
\begin{equation}
M_{HII}(min) = \int _{z_{IF}}^\infty 2dz \int_0^{r_g} m_Hn_e(r,z) 2\pi rdr,
\end{equation}
where $m_H$ is the mass of the ionized gas per electron ($\sim 2 \times
10^{-24}$ gm).  The $z$ integral can be approximated as $2H(r)\equiv
2r_g\left({r \over r_g}\right)^{3/2}$ with the electron density fixed at the
density at the ionization front $n_e(r,z_{IF})$.  The density $n_e(r,z_{IF})$
falls as $r^{-3/2}$ for $r<r_g$.  Therefore, for $r<r_g$ the mass is mostly at
$r_g$.  Performing the integral, we obtain
\begin{equation}
M_{HII}(min)\simeq 2 \pi r_g^3 m_H n_e(r_g, z_{IF}) \simeq 10^{-10}\left({\Phi _{EUV} 
\over {10^{41} \  {\rm s^{-1}}}}\right)^{1/2}\left({{M_*} \over {1\ {\rm M}_\odot}}\right)^{3/2} 
\ {\rm M}_\odot .
\end{equation}
Thus, assuming $n_e(r_g) < n_{ecr,t}$, 
an extremely small gas mass, of order $10^{-7}$ Jupiter masses, 
 inside of $r_g$ will provide the luminosities given
by Eq. (8), using the values of $C_t$ given in Table 1.   These lines, then,
are very sensitive diagnostics of the presence of trace amounts of gas
at radii of order 1-10 AU in disks.
If there is less mass than $M_{HII}(min)$ at $r_g$, then the resulting
luminosities will be reduced by a factor $M_{HII}/M_{HII}(min)$.

We now address whether dust extinction is important in the HII surface region.
Since $H(r) \propto r^{3/2}$, the $10^4$ K surface of the disk is flared.  Most of the emission
comes from $r_g$ or beyond as shown above.  Using Eqs. (11 and 12), 
the attenuating column at $r_g$ is roughly 
\begin{equation}
N_{att} \sim n_e(r_g, z_{IF}) \times r_g \sim 5 \times 10^{18} \left({{\Phi _{EUV}}
\over {10^{41} \ {\rm s}^{-1}}}\right)
^{1/2} \left( {{ 1 \ {\rm M}_\odot}\over
{M_*}}\right)^{1/2} \ {\rm cm}^{-2}.
\end{equation}
The dust at the surface of the disk is expected to have less opacity than
interstellar dust, so that the column for EUV optical depth unity  is
greater than 10$^{21}$ cm$^{-2}$.  The above equation shows that dust will
not be important until $\Phi _{EUV} > 10^{46}$ s$^{-1}$ for a 1 M$_\odot$
star.  Low mass stars do not produce such high EUV luminosities.  Therefore,
our neglect of dust opacity in the HII surface is justified in disks around low mass
stars.

Finally, we estimate the fraction $f_{EUV}$ of the ionizing photons emitted by
the star that are absorbed by the disk. 
 Most of the absorption (and consequent emission,
as shown above) comes from $r \sim r_g$ or beyond,
and here $H \sim r$.  As we will show below in our numerical disk models,
the underlying neutral disk also has considerable height.  At $r= 10$ AU, $z_{IF}\simeq 7.5$ AU
(see \S 4.2).
Thus, the disk is opaque to EUV photons from the midplane to an angle from midplane of about 40 degrees.  
In addition, the recombining hydrogen in the ionized gas above 40 degrees also absorbs EUV photons.
Therefore, we
estimate that the fraction $f_{EUV}$ of EUV photons absorbed by the disk is about
0.7.  A detailed hydrodynamical study is needed to more accurately determine
this fraction.

As our prime examples for this analytic analysis, consider specifically the 
case of the [NeII] 12.8 $\mu$m, [NeIII] 15 $\mu$m, and 
[ArII] 7 $\mu$m  luminosity emerging from a young disk.  We choose these lines
because [NeII] and [NeIII] have been observed, and [ArII] is predicted to
be the brightest of the unobserved lines (see Table 1).  In addition, they
all have high critical densities so that $n_e(r_g,z_{IF})<n_{ecr}$ for these
lines as long as $\Phi_{EUV} < 10^{42} -10^{43}$ s$^{-1}$.  
 Assuming this condition is met  
 and substituting $f_{EUV}= 0.7$ and the atomic constants into Eq. (8), we obtain
\begin{equation}
L_{[NeII]} \simeq 3.8 \times 10^{-6} f(Ne^{+})\left( {{\Phi _{EUV}}\over {10^{41} \ {\rm s}
^{-1}}}\right) \ {\rm L}_\odot .
\end{equation}
\begin{equation}
L_{[NeIII]} \simeq 6.4 \times 10^{-6} f(Ne^{++})\left( {{\Phi _{EUV}}\over {10^{41} \ {\rm s}
^{-1}}}\right) \ {\rm L}_\odot .
\end{equation}
\begin{equation}
L_{[ArII]} \simeq 3.2 \times 10^{-6} f(Ar^+)\left( {{\Phi _{EUV}}\over {10^{41} \ {\rm s}
^{-1}}}\right) \ {\rm L}_\odot .
\end{equation}
Recall that $f(Ne^+)$ is the fraction of neon in the singly ionized state
in the region near $r_g$ which produces most of the emission.
Luminosities greater than about $10^{-7}$ L$_\odot$ are detectable from
nearby ($<100$ pc) sources by the {\it Spitzer Space Telescope}, as long
as the line to continuum ratio is sufficiently large to enable detection of the
line above the bright continuum. 

{\it The effect on IR luminosity caused by holes in disks}.   The above
analytic results apply for a disk that extends inwards to  $r\lta r_g \sim $ 7 AU 
from the star.  However, disks have been observed with inner holes, devoid of dust,
that extend to $r_i > r_g$ (e.g.  Najita et al. 2007, Salyk et al 2009).
 Regardless of the cause of these holes,  if gas is absent inside of $r_i$ and $r_i > r_g$,  the
disk  vigorously photoevaporates at $r_i$, a process which  evaporates the
disk from inside out (Alexander et al 2006a,b).   Alexander et al show that the flux of
EUV photons striking the inner wall of the disk creates a thin (thickness $\simeq H$,
the scale height of the neutral disk at $r_i$) ionized layer.
The Str\"omgren
condition gives the electron density in the layer:
\begin{equation}
F_{EUV} \equiv {\Phi _{EUV} \over {4 \pi r_i^2}} = \alpha _{r,H} n_e^2 H
\end{equation}
Assuming $H\simeq 0.1 r_i$ (Alexander et al. 2006a,b), we obtain
\begin{equation}
n_e(r_i) \simeq  2.5 \times 10^5 \left({\Phi _{EUV} \over {10^{41} \ {\rm s^{-1}}}}\right)^{1/2}
\left({10\ {\rm AU} \over r_i}\right)^{3/2} \ {\rm cm}^{-3}.
\end{equation}
Note that the electron density decreases as the inner hole size increases.  If
$r_i >>r_g$, this leads to an increase in the luminosity of low critical density lines
with respect to high critical density lines [see Eq. (8)].   However, for lines
whose critical densities are larger than the electron density at $r_g$, the
presence of a hole of size $r_i >>r_g$ does not appreciably affect the IR line
luminosity.  Essentially, this arises because the IR line luminosity is proportional to the 
number of EUV photons absorbed, $f_{EUV}\Phi _{EUV}$, and this remains constant, with  $f_{EUV} \sim 0.7$, regardless of $r_i$. Therefore, the IR line luminosity tracks $\Phi _{EUV}$ as presented in Eq. (8). 

\subsection{ Infrared Hydrogen Recombination Lines from the HII Surface}
  The infrared
hydrogen recombination lines can be analytically determined by noting that
the luminosity in a given line produced by the transition $n_u$ to $n_l$ is given
\begin{equation}
L_{ul} =  \alpha _{ul} \Delta E _{ul} \int ^{\infty} _{z_{IF}} 
dz \int ^{r_o} _{r_i} 2 \pi r n_e^2 dr,
\end{equation}
where $\alpha _{ul}$ is the rate coefficient for recombinations through
the $n_u - n_l$ transition, and $\Delta E_{ul}$ is the energy of the
photon produced in this transition.  Clearly, the hydrogen recombination lines
also count EUV photons (see Eq. 5) and could be used to measure $\Phi _{EUV}$.
However, hydrogen recombination produces weak IR lines compared to the
fine structure lines such as [NeII] if the electron density $n_e$ is less than the
critical density  $n_{ecr}$ of the fine
structure transition, as can be seen by taking the ratio
of the predicted line luminosities
\begin{equation}
{{L_{ul}}\over {L(NeII)}} = {{ \alpha _{ul} \Delta 
E _{ul}} \over { \gamma _{[NeII]} \Delta E_{NeII} x_{Ne} f(Ne^+)}} .
\end{equation}
The hydrogen recombination lines we are most interested in are the 7-6 (Humphreys
$\alpha$) and 6-5 (Pfund $\alpha$) at  wavelengths of 12.37 and 7.46 $\mu$m
respectively.  These two lines have been reported observed toward stars with
disks (Pascucci et al. 2007, Ratzka et al. 2007).
  Substituting the atomic constants for these transitions
we obtain predicted ratios for the EUV-induced surface HII layer:
\begin{equation}
 {L_{76}\over L(NeII)} = 0.008
\end{equation}
and 
\begin{equation}
 {L_{65}\over L(NeII)} = 0.02
\end{equation}
The observed ratios are close to unity!  The predicted ratios are small
because of the low ratio of the radiative recombination rate $\alpha _{ul}$
of hydrogen to the electronic collisional excitation rate of [NeII] $\gamma _{[NeII]}$.
Thus, we predict that these IR hydrogen recombination lines 
 must originate from another source if the [NeII]
originates from the HII region surface of the disk.  One place which would
provide copious recombination line emission without producing even more
[NeII] emission would be very high electron density regions, $n_e >> n_{ecr, [NeII]}$.
In these regions the [NeII] is suppressed relative to the hydrogen recombination lines due
to the collisional de-excitation of the upper level of the [NeII] transition.
Therefore, these observed recombination lines are possibly
 produced in very dense plasma very close to the star, in the stellar
chromosphere, the accretion shock, or an internal wind shock if it is both high 
speed ($v_s \gta 100$ km s$^{-1}$ so that it produces ionized hydrogen) and occurs
so close to the wind origin ($\lta 1$ AU) that the postshock density is high
enough to suppress [NeII] relative to the recombination lines. In any of these cases, the prediction 
is that the H recombination lines will be much broader ($\ga100$ km s$^{-1}$)
than the [NeII] lines ($\sim 20$ km s$^{-1}$) in face-on disks. 

\subsection {IR Fine Structure Lines from the X-ray Heated and Ionized Subsurface
Layer}
  Glassgold et al (2007) and Meijerink et al (2008) have presented
models of the [NeII] 12.8 $\mu$m emission  and emission from other
lines, such as [NeIII], [OI], [SIII] and [SIV], produced in the X-ray
heated layer that lies just below the ionization front created by EUV
photons.    This layer is predominately neutral, $x_e \sim 0.001 - 0.1$, depending on
$r$ and $z$, but with a higher 
ratio of Ne$^+$/Ne.  Typically, the [NeII] emitting layer has $T \sim 1000 -4000$ K.
We also in \S 4 present numerical results from our models  of the X-ray-induced
fine structure emission, and in \S 5 discuss differences between our models
and those of these authors.  Here, we present a simple analytic
estimate of the strengths of the fine structure transitions in X-ray-illuminated
regions.  These estimates are more approximate than those presented above for the EUV-dominated
regions because of the uncertainties in estimating the 
gas temperature in this
mostly neutral gas  illuminated by a spectrum of X-ray photons.
Nevertheless, they provide insight into the X-ray process, and into the
relative strengths of X-ray induced fine structure emission in the X-ray layer  as opposed to that
produced by EUV photons in the surface EUV layer.

The simplest derivation arises if we assume our "hard" X-ray spectrum,
dominated by 1-2 keV photons.   These photons are sufficiently energetic
to ionize the K shell of Ne, and the ionization of Ne is dominated by direct
X-ray photoabsorption, and not by collisions with secondary electrons.
If we make the assumption in the X-ray layer that the atomic Ne absorbs
a fraction $f_{Ne}^X$ of all $\sim 1$ keV X-rays, and that Ne$^+$ radiatively
recombines with electrons,  and we assume that $n_e < n_{ecr,[NeII]}$, 
we obtain in a manner completely analogous
to the EUV layer's Eq. (5-7):
\begin{equation}
L_{[NeII]}^X = {{\gamma _{[NeII]}^X \Delta E_{[NeII]} f_{Ne}^X f_X \Phi _X} \over
{2\alpha _{r,Ne}}}.
\end{equation}
Here, $\gamma _{[NeII]}^X$ is the collisional rate coefficient for [NeII] by electrons
in the X-ray layer (that is, it is only different from $\gamma _{[NeII]}$ in Eq. (7)
because the X-ray layer is cooler than the EUV layer), $\Phi _X$ is the $\sim 1$ keV
X-ray photon luminosity of the central source, $f_X$ is the fraction of X-rays
absorbed by the disk in the X-ray layer, and $\alpha _{r,Ne}$ is the
recombination rate coefficient of Ne$^+$ with electrons in the X-ray layer.
The fraction of $\sim 1$ keV photons absorbed by neon, $f_{Ne}^X$, is
approximately the neon cross section at 1 keV divided by the total
X-ray absorption cross section at 1 keV; using Wilms, Allen, \& McCray (2000),
we obtain $f_{Ne}^X \simeq 0.21$.   We see that $L_{[NeII]}^X$ scales linearly with
$\Phi _X$, just as $L_{[NeII]}$ scales linearly with $\Phi _{EUV}$ in the EUV layer.
The ratio of the [NeII] luminosity from the EUV layer to that in the X-ray layer is
then given
\begin{equation}
{L_{[NeII]} \over L_{[NeII]}^X} = \left({\gamma _{[NeII]}\over \gamma _{[NeII]}^X}\right) 
\left({f_{EUV} \over {f_{Ne}^X f_X}}\right)
\left({\alpha _{r,Ne} \over \alpha _{r,H}}\right)
\left({\Phi _{EUV} \over \Phi _X}\right) x_{Ne}f(Ne^+)
\end{equation}
where $f(Ne^+)$ is the fraction of neon that is singly ionized in the EUV layer.
We take $T\sim 10^4$ K for the EUV layer and $T_X\sim 2000$ K for the X-ray layer
to estimate the recombination coefficients, and assuming that $f_X$ is approximately the
height of the layer which becomes optically thin to 1 keV X-rays from the star (roughly
$N\sim 10^{21}$ cm$^{-2}$, or a column of about $10^{22}$ to the star) divided by $r$,
or $f_X \sim 0.25$ (see Fig. 5). 
The EUV layer is more flared, hence $f_{EUV} \sim 0.7$.  In the EUV layer 
$f(Ne^+) \simeq 1$.  Substituting into Eq. (29), we obtain
\begin{equation}
{L_{[NeII]} \over L_{[NeII]}^X} \simeq 2 \times 10^{-3} e^{{1100\ {\rm K} \over T_X}} \left(
{\Phi _{EUV} \over \Phi _X}\right).
\end{equation}
The 1 keV X-ray photon luminosity $\Phi _X$ from a typical source is of order
$10^{39}$ photons s$^{-1}$.  The EUV luminosity $\Phi _{EUV}$ is generally of order
$10^{41}$ photons s$^{-1}$.   Therefore, assuming the 1 keV X-rays are absorbed
in regions with $T_X > 1000$ K, the [NeII] luminosity is expected to be marginally
dominated by emission from the X-ray layer as opposed to the EUV layer.   We
show below in our numerical work that  $L_{[NeII]}/ L_{[NeII]}^X \sim 0.6$ when
$\Phi _{EUV} = 10^{41}$ s$^{-1}$ and $\Phi _X \simeq 10^{39}$ s$^{-1}$ and when the EUV
spectrum is such to produce more Ne$^+$ than Ne$^{++}$ in the EUV layer; this result
agrees with Eq. (30).  Note that $\Phi _{EUV} = 10^{41}$ s$^{-1}$ and $\Phi _X \simeq 10^{39}$ s$^{-1}$
corresponds to $L_{EUV} \simeq L_X$.  In other words, if the central star emits the
same EUV and X-ray luminosity, and the EUV has a soft spectrum which produces
more [NeII] than [NeIII], there will be roughly 2 times more [NeII] luminosity arising from the X-ray
layer than from the EUV layer.  As we will show in \S 4, where we present results
from our detailed numerical models,  this conclusion that {\it X-rays are more efficient
at producing [NeII] emission} does
not depend strongly on the X-ray spectrum for reasonable choices of the spectrum.
If we adopt a softer spectrum, the ionization of Ne is dominated by secondary electrons
because most of the X-rays are absorbed by He, C or O.    The gas is also hotter because
there is more heating per unit volume due to the higher cross sections for softer X-rays.
The net effect is that the [NeII] luminosity does not change much for fixed X-ray luminosity,
even as we vary the X-ray spectrum.
In the case of the "hard" X-ray spectrum, the reason X-rays are somewhat more dominant
than the EUV  is because for high temperature ($T>1000$ K) gas such as
exist in both the HII region and the X-ray heated region, the luminosity in the line depends
mainly on the number of Ne$^+$ ions times the electron density.   In the HII region, the 
vast number of EUV photons
are used ionizing H, an extremely small fraction of the photons are used
ionizing Ne, and therefore the number of Ne$^+$ ions  times the electron density is small (the
$x_{Ne}f(Ne^+)$ factor in Eq. (29);   however, we explicitly include the relatively
large fraction, $f_{Ne}^X= 0.21$ of 1 keV X-ray photons that directly ionize Ne
and lead to a large product of  electron density times Ne$^+$ ions in Eq. (29).
Therefore, no $x_{Ne}f(Ne^+)$ term appears in the denominator. Another way of
understanding this result is that, although the EUV layer is completely ionized with
$f(Ne^+) \sim 1$ and $x_e \sim 1$, the EUV layer has much less column of Ne$^+$
because H and He rapidly absorb the EUV photons; the penetrating X-rays partially
ionize a much larger column.

\subsection{Shock Origin of Ionized Infrared Fine Structure Lines}
Lahuis et al (2007) and van Boekel et al (2009) discuss the possible origin
of [NeII] emission from shocks generated by protostellar outflows.   In order
for a shock to produce significant [NeII] emission, the shock must ionize most
of the preshock gas in order to produce high quantities of Ne$^+$ and electrons.
The fraction of preshock gas that is ionized by the shock is a very sensitive function
of the shock speed $v_s$ (e.g, Hollenbach \& McKee 1989, hereafter HM89).
HM89 show that $v_s \gta 100$ km s$^{-1}$ is required to ionize most of the H
and Ne and that the [NeII] emission rises very sharply with $v_s$ and then
plateaus above $v_s \gta 100$ km s$^{-1}$.  This suggests that any possible shock
must originate from internal shocks in the protostellar wind (which has terminal
speeds of $\sim 200$ km s$^{-1}$) or from the protostellar
wind overtaking much slower moving outflow material.  It is unlikely to originate from
the shock produced by the wind striking the disk, since this shock is so oblique that
the (normal) shock speeds are typically $\lta 20$ km s$^{-1}$ (Matsuyama, Johnstone
\& Hollenbach 2009).   The total emission per unit area $F_t$ from the postshock region
of a radiative shock is given:
\begin{equation}
F_t = {1\over 2} m_H n_0 v_s^3,
\end{equation}
where $m_H$ is the mass per hydrogen nucleus and $n_0$ is the hydrogen nucleus
number density of the gas.  The numerical results of HM89 can be approximated for
the emission per unit area of [NeII] for shocks with $v_s \ga 100$ km s$^{_1}$:
\begin{equation}
F_{[NeII]} \simeq \left( {{5 \times 10^{-3}}\over {1 + {n_0 \over {10^4 \ {\rm cm}^{-3}}}}}\right) F_t,
\end{equation}
where the dependence on density at high density arises because of collisional
de-excitation of the upper state of [NeII] in the postshock gas. Assume that a fraction
$f_{sh}$ of the protostellar wind shocks at speeds $v_s \sim 100$ km s$^{-1}$ with
preshock density $n_0$.  It follows that the [NeII] shock luminosity is
\begin{equation}
L_{[NeII]}^{sh}= {1\over 2} f_{sh} \dot M_w v_s^2 \left({5\times 10^{-3} \over {1 + {n_0 \over {10^4 \ 
{\rm cm}^{-3}}}}}\right) \simeq \left({4\times 10^{-5} \over {1 + {n_0 \over {10^4 \ {\rm cm}^{-3}}}}}\right)
f_{sh} \dot M_{-8} \ {\rm L}_\odot,
\end{equation}
where $\dot M_{-8} = \dot M_w/10^{-8}$ M$_\odot$ yr$^{-1}$ and $v_s \gta 100$ km s$^{-1}$.
Therefore, if the protostellar wind mass loss rate $\dot M_w \gta 10^{-9}$ M$_\odot$ yr$^{-1}$,
$v_s \gta 100$ km s$^{-1}$, $f_{sh} \sim 1$, and $n_0 \lta 10^4$ cm$^{-3}$, the
[NeII] luminosity produced in internal wind shocks may be comparable to or greater than
the luminosity produced in the EUV or X-ray layer of the disk.

 Van Boekel et al (2009) argue this may be the case in T Tau S.  Here, the observed [NeII] luminosity is
 $L_{[NeII]} \sim 10^{-3}$ L$_\odot$.  From the above, this would require, for example, $f_{sh} \sim 1$,
 $\dot M_w \sim 2.5 \times 10^{-7}$ M$_\odot$ yr$^{-1}$, and $n_0 \lta 10^4$ cm$^{-3}$.  The preshock
 density can be estimated with knowledge of the distance of the shock from the star, $r_{sh}$, and
 $\dot M_w$. The preshock density (the density in the wind at $r_{sh}$) is given
 \begin{equation}
 n_0 = 2.5 \times 10^3 \dot M_{-8} r_{15}^{-2} f_\Omega ^{-1} \ \ {\rm cm}^{-3},
 \end{equation}
 where $r_{15}=r_{sh}/10^{15}$ cm and $f_\Omega$ is the fraction of 4$\pi$ sterradians into which
 the protostellar wind is collimated.   Van Boekel et al (2009) measure an extent of the emission
 from T Tau S of approximately 160 AU.  If we assume $r_{sh}= 160$ AU and $f_\Omega = 1$,
 presumably upper limits, we obtain a lower limit to  $n_0 \gta 1.5 \times 10^4$ cm$^{-3}$.   Note that for $\dot M_w
 > 2.5 \times 10^{-7}$ M$_\odot$ yr$^{-1}$ and for our specific assumptions on $r_{sh}$ and $f_\Omega$,
 the preshock density is so high that the [NeII] luminosity is independent of $\dot M_w$; the 
 luminosity from the shock saturates once the
 emitting [NeII] in the postshock gas is in LTE.   Therefore, although it pushes parameters a bit
 uncomfortably, if T Tau S has protostellar mass loss rates $\gta 2.5 \times 10^{-7}$ M$_\odot$ yr$^{-1}$,
 it is possible that internal wind shocks produce the observed [NeII] luminosity.  Note that the ``dynamical
 time", $v_s/r_{sh}\sim 10$ yrs, is marginally consistent with the observations of no significant time
 dependence since 1998 (van Boekel et al 2009).
    
     More recently, G\"udel et al (2009) have assembled [NeII] data from a large number of sources
  and have plotted $L_{[NeII]}$ versus $\dot M _{acc}$.   For low values of $\dot M_{acc} \lta
  10^{-8}$ M$_\odot$ yr$^{-1}$ the [NeII] luminosity is nearly independent of $\dot M_{acc}$,
  and is typically of order $3 \times 10^{-6}$ L$_\odot$.
  However, for higher mass accretion rates, and in particular, {for all the sources with known
  outflows or jets}, $L_{[NeII]}$ increases with increasing $\dot M _{acc}$ (arguably linearly, but
  with a lot of scatter).  We present these observational results from G\"udel et al and compare
  them to Eq. (33) in \S 4.4.   The correlation of $L_{[NeII]}$ with $\dot M_{acc}$ 
   suggests either that the higher luminosity ($L_{[NeII]} \sim 10^{-5} -
  10^{-3}$ L$_\odot$) sources may originate in protostellar shocks or from EUV or soft X-rays produced
  by the accretion of disk material onto the star.  However, in the latter case, these photons must
  penetrate the disk wind, which seems unlikely.
 
 \subsection{[OI] 6300 \AA \ Emission from Young Stars with Disks}
 [OI] 6300 \AA \ emission is often observed in young low-mass stars with disks and outflows (e.g., Hartigan
 et al 1995).  Two velocity components are seen: a high velocity component "HVC" and a low velocity
 component ``LVC".   Hartigan et al argue that the HVC comes from shocks in the protostellar wind,
 similar to our discussion above of internal shocks.  The typical velocity of this component is $\sim 100
 - 200$ km s$^{-1}$, and the [OI] 6300 \AA \ luminosity is $\sim 10^{-6} - 10^{-2}$ L$_\odot$.   HM89
 show that for $v_s \sim 100$ km s$^{-1}$ the [OI] 6300 \AA \ emission from the shock is about 10 times
 more luminous than the [NeII] 12.8 $\mu$m emission for $n_0 \lta 10^4$ cm$^{-3}$, with even higher ratios
 at $n_0 > 10^4$ cm$^{-3}$ because [OI] 6300 \AA \ does not collisionally de-excite as readily as [NeII]
 12.8 $\mu$m.   Therefore, in agreement with Hartigan et al, we find that mass outflow rates of $\gta 10^{-7}$
 M$_\odot$ yr$^{-1}$ can produce the most luminous [OI] 6300 \AA \  HVC sources (see Eq. 33).
 
 Hartigan et al (1995) attribute the LVC to [OI] emission emanating from the disk surface, probably in a relatively
 slow outflow since the emission is observed to be slightly blueshifted ($\sim - 5$ km s$^{-1}$ with great 
 dispersion).  However, there are red and blue wings extending to $\pm 60$ km s$^{-1}$ in the LVC,
 presumably due to a combination of Keplerian rotation and outflow.  The [OI] 6300 \AA \ luminosity
 in the LVC ranges from $\sim 10^{-6}$ to $\sim 10^{-3}$ L$_\odot$, with "typical" values $\sim 10^{-4}$
  L$_\odot$ (Hartigan et al 1995).  The exact origin of the LVC [OI] emission, and its heating source, remains a mystery.
 In Section 4 below we present our numerical model results for [OI] emission from the EUV layer and
 the X-ray layer.  In agreement with previous work by Font et al (2004), we find that the EUV layer can only provide
 [OI] 6300 \AA \ luminosities $\lta 10^{-6}$ L$_\odot$.  Meijerink et al (2008) were able to produce [OI]
 luminosities as high as $5\times 10^{-5}$ L$_\odot$ in their models of the X-ray layer.   Therefore, they found
 it very difficult to explain the most luminous [OI] LVC sources, but were able to produce luminosities in accordance
 with many of the observations.     However, our more detailed models with a similar X-ray spectrum
 as that assumed by Meijerink et al produce lower [OI] luminosities, primarily because 
 we obtain lower gas temperatures in the X-ray heated layer (see below).   However, if we use
 a softer X-ray spectrum such as the one proposed by Ercolano et al (2009), we do obtain
  luminosities of order  $10^{-4}$ L$_\odot$.  In summary, 
 it appears that emission from the EUV and X-ray layers can only explain the
 lower and typical luminosity LVCs, but not the highest luminosity LVCs.
 
 It is instructive to estimate what physical conditions are required to produce [OI] luminosities in the LVC
 as high as $10^{-4} - 10^{-3}$ L$_\odot$.  Consider a layer on the disk surface of thickness $\ell$, 
 temperature $T$, and extending to radius $r_o$ from which most of the [OI] emanates.  This top and bottom
 layer of the disk has hydrogen nucleus density $n$ and vertical column $N$.   Because the [OI] 6300
 \AA \ transition is $\Delta E/k=23,000$ K above the ground state, we require $T\gta$ several thousand
 degrees K for significant emission.  The emerging [OI] 6300 \AA \ luminosity produced by the surface layers is then
 given
 \begin{equation}
 L_{[OI]}= \pi r_o^2n_e n(O)\gamma _{[OI]} \ell,
 \end{equation}
 where $\gamma _{[OI]}$ is the collisional excitation rate for electrons on atomic oxygen, $n(O)$ is the density
 of atomic oxygen, and $n_e$ is the electron density.  We account here for both the 
 top and bottom of the disk, but recall that 1/2 of the luminosity (that directed to the midplane) is absorbed by
 the optically thick disk.  Eq. (35) assumes $n_e$ 
 to be less than the critical density $n_{ecr}$.  (HM89 give $n_{ecr} \sim 10^6$ cm$^{-3}$ so
 $n_e < n_{ecr}$ is generally satisfied).   HM89 give $\gamma _{[OI]} =8.5 \times 10^{-9} T_4^{0.57} e^{-2.3/T_4}$
 cm$^3$ s$^{-1}$, with $T_4 = T/10^4$ K.  Oxygen rapidly charge exchanges with hydrogen and therefore
 at high temperatures ($T>>200$ K) O$^+$/O = H$^+$/H.  Therefore, $n_e=x_e n$ and $n(O)=
 x(H)n_O$, where $n_O \simeq 3 \times 10^{-4}n$ is the gas phase density  of oxygen in both O and
 O$^+$ and $x(H)$ is the abundance of atomic hydrogen.  It follows that
 \begin{equation}
 L_{[OI]} \simeq 6.5 \times 10^{-4} x_e x(H) n_5 r_{14}^2 N_{20} T_4^{0.57} e^{-2.3/T_4} \ {\rm L}_\odot,
 \end{equation}
 where $r_{14}= r_o/10^{14}$ cm, $n_5 = n/10^5$ cm$^{-3}$, and $N_{20}= N/10^{20}$ cm$^{-2}$. 
 This analytic exercise shows that to produce $L_{[OI]}\sim 10^{-4} - 10^{3}$ L$_\odot$
in the LVC requires, for example,  surface layers with $n \sim 10^5$ cm$^{-3}$, $N\sim 10^{20}$ cm$^{-2}$,
$r_o \sim 60$ AU, $T \sim 10^4$ K, and $x_e \sim 0.5$.
In the EUV layer $x_e \sim 1$, $T_4 \sim 1$, and $r_{14}^2 \sim 1$ (recall most of emission arises from
 $r \sim r_g$).  Therefore, the [OI] luminosity from the EUV layer is approximately
\begin{equation}
L_{[OI]}^{EUV} \simeq 6.5 \times 10^{-5} x(H) r_{14}^2 n_5 N_{20} \  {\rm L}_\odot.
\end{equation}
This equation shows the difficulty in producing the observed [OI] emission from the EUV layer: the fraction of
neutral gas $x(H)$ is very low in the EUV layer, or equivalently, the fraction of atomic oxygen is
very low.   In addition, the emission mostly arises from $r\sim r_g \sim 10^{14}$ cm, which is not large, and
$n_5 N_{20}$ rarely exceeds unity.

On the other hand, in the X-ray layer the temperature is of order $T_4 \sim 0.1 - 0.4$  and  $x(H)\sim 1$ so that
\begin{equation}
 L_{[OI]}^X \simeq 6.5 \times 10^{-4} x_e  n_5 r_{14}^2 N_{20} T_4^{0.57} e^{-2.3/T_4} \ {\rm L}_\odot.
 \end{equation}
The X-ray layer has $x_e \lta 0.1$, $r_{14}^2  \lta 10$, $n_5 \lta 100$ and $N_{20} \lta 10$.   Even inserting these upper
limits, we find that the [OI] luminosity is at most $10^{-3}$ L$_\odot$ if $T= 4000$ K and $3 \times 10^{-6}$
L$_\odot$ if $T= 2000$ K.  Therefore, the [OI] luminosity is extremely sensitive to the gas temperature (and
its variation in $r$ and $z$) in the X-ray layer.  The Meijerink et al (2008) model has high 
temperatures, $\sim 3000-4000$ K,
in the X-ray layer out to 10-20 AU in their case with relatively high X-ray luminosity of $2 \times 10^{31}$ erg s$^{-1}$.
For this case, they therefore find $ L_{[OI]}^X \sim 5\times 10^{-5}$ L$_\odot$.   Our model for the same case, 
however, gets temperatures in the X-ray layer in the range $1500-2500$ K, and therefore we get an [OI] luminosity
of only $\sim 5 \times 10^{-7}$  L$_\odot$ (see \S 4.4 below).    However, keeping the X-ray luminosity the same
but assuming a much softer power law spectrum [$L(E) \propto E^{-1}$ from 0.1 to 2 keV], we do find
that the X-ray heated gas becomes hotter and the [OI] luminosities approach $ 10^{-4}$ L$_\odot$. We discuss
 in \S 4.3 the reasons for the differences in temperature in our model relative to that of Meijerink et al.  We conclude that it is unlikely that the X-ray layer can provide 
the highest  [OI] luminosities observed in some LVC sources, but that soft X-ray sources can
produce the typical luminosities.

  The ``transition layer" ($x_e \sim 0.5$) between
the fully ionized EUV layer and the partially ionized X-ray layer is also unlikely to produce 
either the highest observed
[OI] luminosities, or even the typical luminosities.   
The density in this layer is similar to the density at the base of the EUV layer,
or $n \sim 10^5 \Phi _{41}^{1/2} $ cm$^{-3}$ at $r=r_g = 7$ AU (see Eq. 12).  However, beyond
$r_g = 7$ AU the density falls as $r^{-5/2}$ so that most of the emission arises from $r \sim 7$ AU.
In addition, the column $N$ of this transition layer is roughly the column for optical depth unity
in EUV photons, or $N\sim 10^{18}$ cm$^{-2}$.  Therefore, the small $r_0$ and low $N$ conspire
to produce only $L_{[OI]} \sim 10^{-6}$ L$_\odot$.

We plan to investigate the possibility that the source of the LVC arises from the shear layer produced
when the protostellar wind strikes the surface of the disk obliquely and sets up an outward moving layer
of shocked wind and entrained disk surface gas (a "shear" layer, see Matsuyama et al 2009), hereafter
MJH09).   MJH09 show that this layer can have columns $N\sim 10^{19}-10^{20}$ cm$^{-2}$ out
to 100 AU.  As noted earlier, the oblique wind shock ($v_s \lta 20$ km s$^{-1}$) is insufficient to ionize 
hydrogen or helium to provide the electrons needed for [OI] excitation.
  Therefore, we require the EUV and soft X-rays to ionize this layer.   This shear layer is likely 
turbulent so that there may be rapid mixing of the bottom (cooler and more neutral layers) with the top shear layers,
perhaps allowing $x_e \sim 0.5$ and $x(H) \sim 0.5$ in the entire layer (the most efficient for producing [OI]) and maintaining a relatively isothermal layer.
The heating would be a combination of shock/turbulent heating plus the heating due to 
photoionization by EUV and soft X-rays.   The density $n$ in the shear layer (MJH09) is approximately
$n \sim 3000 \dot M_{-8} r_{15}^{-2}$ cm$^{-3}$.  Note that we cannot allow $\dot M_{-8}$ to exceed
unity or the EUV and soft X-rays will not be able to penetrate the base of the wind to heat and ionize the
shear layer.  Therefore, although the shear layer may provide sufficient column $N$, electron fraction
$x_e$, and temperature $T$, it appears unlikely to produce sufficient $nr^2$ to give the observed
[OI] luminosities in the more luminous sources.   Further work is needed to confirm this rough
argument.

We conclude that the origin of the very luminous LVC [OI] emission is not from the EUV layer, the X-ray
layer, or the transition layer. The typical LVC [OI] emission, however, may be produced by soft X-rays.
The LVCs are also unlikely to originate from the shear layer set up by the impact of the protostellar
wind.  Perhaps a model that invokes ambipolar diffusion as a heating source, such as those that Safier 
(1993a,b) has constructed for the HVC, might be applicable for the most luminous LVCs.   
However, the EUV layer
is capable of producing the lowest luminosity sources, and the X-ray layer may produce the
typical luminosity, so we proceed with detailed numerical studies of the [OI] luminosity from
these layers in \S 4 below.

\section{NUMERICAL MODEL AND RESULTS}
\subsection{The EUV Surface Layer and the Underlying X-ray Layer}

Gorti \& Hollenbach (2008) describe the numerical code which we use to 
calculate, self-consistently, the gas temperature, gas density, and chemical structure
of the predominantly neutral gas in the disk.  To summarize, the code includes $\sim$ 600 reactions
among 84 species, gas heating by a number of mechanisms including FUV grain/PAH photoelectric
heating and the heating caused by X-ray ionization ionization of the gas, and cooling not only from collisional
excitation of the species followed by radiative decay, but also from gas-grain collisions
when the dust is colder than the gas.  In some instances, for example deep in the disk below
the surface layers, the dust is warmer than the gas in which case gas-grain collisions can
be an important heating source for the gas.  The vertical structure of the disk is calculated
self-consistently by using the computed gas temperature and density to calculate the thermal
pressure, and then balancing thermal pressure gradients with the vertical (downward) 
gravitational force from the central star.

For this paper we consider  the ``chemistry" of the fully ionized (HII) surface region, photoionized
by the EUV radiation field from the star.  By ``chemistry" we mean the computation of the
different ionic states of a given element by balancing photoionization with electronic recombination
and charge exchange reactions.  Photoionization rates are computed using cross sections from Verner et al. (1996).  Recombination rates are taken from 
Aldrovandi \& Pequignot (1973), Shull \& van Steenberg (1982) and
Arnaud \& Rothenflug (1985).  Charge exchange
rates are from Kingdon \& Ferland (1996).  We assume a gas temperature of $10^4$ K and do not perform
thermal balance calculations in the HII region.  In short, our code 
for the surface of the disk is an HII region code where we assume an EUV spectrum 
from the central star and then compute the abundances of, for example, Ne$^+$,
Ne$^{++}$ and Ne$^{+++}$  at each point $r,z$ in the surface HII region.  At each point $r,z$
we compute both the direct EUV flux from the star and the diffuse EUV field caused by
recombinations to the ground state of atomic hydrogen in the surface HII region.
We use the method described by Hollenbach et al (1994) and utilized by Font et al (2004)
to do both these computations.  The code finds the electron density $n_e(r, z_{IF})$ at
the base of the surface HII region (in other words, at the ionization front separating the 
ionized surface from the predominantly neutral disk below).  The thermal pressure at the
IF is then $P_{IF}\simeq 2 n_e(r, z_{IF})kT_{II}$, where $T_{II}=10^4$ K and the factor of 2 includes
the pressure from protons, He$^+$ and electrons.  This pressure then determines the height $z_{IF}$
where the thermal pressure in the predominantly neutral gas has dropped from its midplane
value to $P_{IF}$.  The parameter $z_{IF}$ is the height of the base of the HII region: above $z_{IF}$
the emission is mostly EUV-induced and we call this region the EUV surface layer; below
$z_{IF}$ the emission is mostly X-ray-induced and we call this region the X-ray layer.\footnote{We
note that FUV may contribute to the heating in the X-ray layer.  However, X-rays produce Ne$^+$
and the electrons necessary for efficient excitation of [NeII] and [OI].  Therefore, it is proper to call
this the "X-ray layer".}

Implicit in our model is the assumption that the EUV luminosity and the X-ray luminosity is generated
close to the stellar surface.    This assumption then allows us to determine the column density of
wind that the EUV and X-ray fluxes must traverse (see \S 2) as well as the angle of incidence of the
EUV and X-ray flux on the flared disk surface.   Since we assume the protostellar disk wind originates
near $r \sim 10^{12}$ cm, our estimate of the attenuation column density at the wind base is valid
as long as the EUV and X-ray source of luminosity originates roughly within this distance from
the stellar surface.  Models of X-ray flares indicate that the X-rays probably arise from flares whose
size ranges from 0.1 to 10 times the stellar radius (e.g., Favata et al 2001, 
Grosso et al 2004, Favata et al 2005, Franciosini et al 2007, Stelzer et al 2007, Getman et al
2008a,b).   Therefore, our estimate of the attenuation column is likely valid.   Similarly, the
line emission that we model usually arises from $r\sim$1-10 AU in the disk, and so the placement
of the EUV or X-ray source within $10^{12}$ cm of the star does not affect our results.   However, in DG
Tau, a soft X-ray source has been imaged and seen to arise about 20 AU from the star, probably
from shocks in a jet (G\"udel et al 2008, Schneider \& Schmitt 2008).    Such a geometry would 
certainly lower the column of attenuating wind,
because of the spherical divergence of the wind.   In addition, the X-ray flux would strike the
disk from above, nearly normal to the surface.   This latter effect, however, will likely not significantly affect
the luminosities in the lines, since as we have shown in \S 3, the line luminosities are really
an emission measure effect, and mainly depend on the fraction of energetic photons that the
disk absorbs.   If the source is  20 AU from the star, roughly half of the energetic photons are
absorbed.  We show below that in the case of a flared disk with a central source of energetic photons,
nearly 0.7 of the photons are absorbed.  Therefore the fraction of photons absorbed is nearly
the same.

Also implicit in our model is that the X-ray luminosity is the mean value of the time variable X-ray
luminosity.    Getman et al (2008a) show that typical decay times of flares is of order hours to days.
In the EUV and X-ray layers where the modeled lines originate, recombination and cooling timescales
are of order 1-10 years.   Thus, the gas generally does not have time to respond to the flares, but settles to
a state given by the mean value of the X-ray luminosity.  

\subsection{The EUV Layer Results}

Figures 1 and 2 show the results of models as we vary the EUV luminosity.  We assume two
forms for the EUV spectrum. The first form (Figure 1) is a relatively hard spectrum; we assume a
power law spectrum $\nu L_\nu =$ constant from 13.6 eV to the X-ray regime ($\sim 0.1$
 keV).  This spectrum is motivated by the fact that $\nu L_\nu $ in the FUV band is
observed to be similar to $\nu L_\nu $ in the X-ray band, and each band typically has
$\nu L_\nu \sim 10^{-3}$ $L_*$, where $ L_*$ is the stellar bolometic luminosity.   On the other hand, the
EUV spectrum is very uncertain.  The Ribas et al (2005) observations of older, but very nearby,
solar mass stars show EUV spectra which can drop rapidly from 13.6 eV to 40 eV, even
though the overall trend from the FUV to the X-ray tends to roughly an $L_\nu \propto \nu ^{-1}$
spectrum.   To simulate a softer spectrum than the first form, we take a blackbody spectrum
with an effective temperature of 30,000 K (Figure 2).  We are further motivated to adopt a softer spectrum
by the observations in one source of the ratio [NeIII] 15 $\mu$m/[NeII] 12.8 $\mu$m \lta 0.06
(Lahuis et al. 2007), and because [NeII] has been detected in more than 25 sources and
none of them show [NeIII].    Our first form of EUV spectrum produces a ratio $>1$!  This either indicates
very little production of [NeII] by  the EUV layer, or that the EUV spectrum is much softer.
We therefore have chosen
a blackbody EUV spectrum that provides ratios in accord with
measured values or upper limits on the ratio.

Figures 1 and 2 show the nearly linear rise in $L_{[NeII]}$ or $L_{[NeIII]}$ with $L_{EUV}$ as
predicted in Eqs. (19) and (20).  The absolute values are also in good agreement with these equations.
 At very high $L_{EUV}$ this linear relationship breaks down for [NeIII],
and  $L_{NeIII}$ begin to saturate because the electron densities in
the dominant emitting regions begin to exceed $n_{ecr,[NeIII]}$.  For our power law spectrum
in Figure 1 we find that by fitting our analytic results to the model 
that $f(Ne^{++}) \sim 0.75$ and  $f(Ne^{+}) \sim 0.25$; i.e., 75\% of the emitting
neon is in Ne$^{++}$ and only 25\% in Ne$^+$.  However,  in Figure 2 we see that a softer EUV
spectrum (blackbody with $T_{eff}= 30,000$ K) will reverse the situation so that [NeII]
dominates.   Another mechanism, not treated here, that would quench [NeIII] and raise
[NeII] would be turbulent mixing of neutral gas into the HII region.  The charge exchange
reaction Ne$^{++}$  + H $\rightarrow$ Ne$^+$ + H$^+$ is very rapid (e.g., Butler \& Dalgarno 1980),
and even a neutral fraction $x(H) \sim 10^{-2}$ would lead to [NeIII]/[NeII] $< 1$. 

Figure 3 shows the results for a number of other fine structure transitions listed
in Table 1.   One sees that [ArII] 7 $\mu$m is the strongest predicted line not yet
observed.   Again, the analytic estimates (Eq. 8 and Table 1) are very good.

Figure 4 shows the radial origin of the EUV-induced emission in the [NeII], [NeIII], [ArII], and
[SIII]  lines.
We use here as a standard case $\Phi _{EUV} = 10^{41}$ s$^{-1}$ ($L_{EUV}=2\times 10^{30}$ 
erg s$^{-1}$) and the
blackbody spectrum,  and $L_X = 2\times 10^{30}$
erg s$^{-1}$ (with our standard X-ray spectrum, see \S 4.3)  although the radial origin is quite insensitive to these parameters.
We see that most of the emission arises from $r \sim r_g \sim 10$ AU, as predicted in
\S 3.  We plot $4 \pi r^2$ times the emergent flux from one side of the disk.   This roughly gives the
luminosity arising from both sides of the disk and from a region extending from $0.5r$ to $1.5r$.
  For dominant ions like Ne$^+$,
the luminosity scales as $n_e^2 H(r) r^2 \propto r^{1/2}$ for $r < r_g$ and $n_e < n_{ecr}$
and as $r^2$ for $n_e > n_{ecr}$. 
 The luminosity scales as $n_e^2 r^3$ for $r > r_g$ and
$n_e < n_{ecr}$, so that here it scales as $r^{-2}$ (see Eq. 13), as seen in [ArII] and [NeII].  For
non-dominant ions such as Ne$^{++}$, the scalings change because the fraction of neon in
Ne$^{++}$ changes with radius.

Hollenbach et al (1994) and Gorti \& Hollenbach (2009) show that significant photoevaporation
flows proceed in the 10$^4$ K gas in the EUV layer at $\gta 1$ AU.   Thus, although 
our models here are static,
the emitting gas is actually rising (and rapidly turning radial, see Font et al 2004) 
at speeds of order  the sound speed (or
10 km s$^{-1}$) from the surface of the disk.  As discussed in Hollenbach et al (1994),
the electron density at the base is not much affected by this flow.   Photoionization
and recombination timescales are sufficiently short that steady state still applies
to the computation of the ionization state of each element in the gas.  Therefore,
we expect our model results on the emitted luminosities in each fine structure line
to be well approximated by the static model solution.  However, the observed line
profiles will be affected by this flow.  For a disk viewed edge on, the lines will be
broadened by the Keplerian rotation but also the radial outward flow.   For a disk
viewed face-on, the lines will be broadened mostly by the radial outward flow,
and since the far side of the disk is obscured, one would expect a blue shift.
Alexander (2008) has recently modeled [NeII] line profiles from photoevaporating
disks.  He predicts broad ($30-40$km s$^{-1}$), double-peaked profiles from edge-on disks due to rotation, and a narrower ($\sim 10$ km s$^{-1}$) profile with a significant blue shift ($5-10$km s$^{-1}$) from face-on disks.  He argues that the observed line widths in TW Hya (Herczeg et al. 2007) are consistent with a photoevaporative wind (see also
Pascucci \& Sterzik 2009).  Resolved [NeII] observations can thus provide a test of EUV photoevaporation models.

Figure 5 shows the vertical origin of the EUV-induced  and X-ray-induced emission at $r=10$ AU for the
standard case.   We have plotted gas temperature $T$, the dust temperature $T_{dust}$,  and the 
hydrogen nucleus density $n$ as a function of the
hydrogen nucleus column $N$ measured from $z=r$ (the putative ``surface" of the disk) downwards.  On the top of the figure we give
the values of $z$ that correspond to those of $N$.  The completely ionized EUV layer extends
to $N\sim 3\times 10^{18}$ cm$^{-2}$, and has electron densities $n_e \sim n \sim 3\times 10^4$ cm$^{-3}$
(see Eq. 12).  The X-ray heated ($T\sim 1000$ K) and ionized layer extends from $N \simeq 3\times 10^{18}$
cm$^{-2}$ to $N\simeq 3\times 10^{20}$ cm$^{-2}$ with hydrogen atom densities $\sim 3 \times 10^6$
cm$^{-3}$. FUV photons also contribute to the heating of this layer.

\subsection{The X-ray Layer}

We model the X-ray layer for both a "soft" X-ray spectrum and a "hard" X-ray spectrum.
Our standard ("hard") X-ray spectrum (Gorti \& Hollenbach 2004,2008) is based on
observed X-ray spectra from young stars, with an attempt to correct for extinction at
the softer energies (Feigelson \& Montmerle 1999).  Our fit to this spectrum is that of
a power law $L_\nu \propto \nu$ from 0.1 keV $< h\nu < 2$ keV, fitting to another power
law $L_\nu \propto \nu ^{-2}$ for $h\nu > 2$ keV.   This spectrum is similar to that adopted
by Glassgold et al (2007) and Meijerink et al (2008).   We also model disks illuminated
by a softer  X-ray spectrum: $L_\nu \propto \nu^{-1}$ for 0.1 keV $< h\nu < 2$ keV, and
$L_\nu \propto \nu ^{-1.75}$ for $h\nu > 2$ keV.\footnote{ We note that our "soft" X-ray 
spectrum has the same power law $L_\nu \propto \nu ^{-1}$ form as our ``hard" EUV spectrum!  
In addition, we note that even
our ``hard" X-ray spectrum includes a contribution from 0.1 keV X-rays.} Ercolano et al 2009 
recently provide evidence that such a spectrum might be 
expected from young, low mass stars.  We note that our soft X-ray spectrum has equal energy flux
in equal logarithmic intervals of photon energy between 0.1 keV and 2 keV; that is, there is as
much energy flux between 0.1 and 0.2 keV as there is between 1 and 2 keV.  We do not consider
here a harder spectrum than our "hard" case, although recently there have been observations
of "superhot" flares (Getman et al 2008a) that indicate significant emission in the 3-8 keV
region of the spectrum.   We do extend our "soft" and "hard" spectra to 10 keV, but there is
insignificant energy flux beyond a few keV.    If there were, then for the same X-ray luminosity
as our two cases, we would obtain less emission in the lines we model in this paper.   The
higher energy photons penetrate more column of gas, depositing less energy per unit volume,
and therefore lead to cooler gas than in our current X-ray layer.  In addition, because of the
increased penetration, the heat is deposited in molecular regions, where the cooling is
enhanced by the molecular transitions.  Therefore the emitting gas is substantially cooler
and most of the X-ray heating energy presumably emerges in molecular rotation lines
of, for example, CO, OH, and H$_2$O or possibly, if grains are abundant, as IR continuum
emission from grains heated by collisions with the warmer gas.  However, we emphasize that
the heating and cooling timescales are long, of order 1-10 years, so that superhot flares that are much
more short-lived than this timescale will not produce a significant effect.

Glassgold et al (2007) first modeled and discussed the physics of the X-ray layer, and our
results are in basic accord with theirs, except, as discussed below, we obtain somewhat
cooler temperatures in the X-ray layer.
Figure 6 shows the vertical structure of the same case plotted in Figure 5 at the same radius, $r= 10$
AU, only the $N$ range is shrunk to emphasize the X-ray layer.  In Figure 6, we plot the electron 
abundance $x_e$ relative to H nuclei, and the fraction of neon in Ne$^+$, $f_{Ne}(Ne^+)$.  The [NeII]
12.8 $\mu$m line luminosity is proportional to $x_e f_{Ne}(Ne^+)$, and very sensitive to $T$ in the
X-ray layer (see Figure 5 and Eq. (28), where the $T$ dependence comes in the collisional
rate coefficient, which is proportional to $e^{-1100/T}$).
Below the EUV layer to a depth $N\sim 3 \times 10^{20}$ cm$^{-2}$, X-rays maintain a 
relatively high fraction of Ne$^+$, $f_{Ne}(Ne^+) \gta 10^{-2}$,
and X-ray ionization of H and He  as well as FUV ionization of C maintain a relatively
high ionization fraction, $x_e \gta 2\times 10^{-4}$ (see Figure 6).
Note that the column attenuating the X-rays in this layer is
the column through the disk {\it to the star}, which is typically $\sim 10 N$, where $N$ is the
vertical column to the disk surface
In the X-ray layer, $10^{19}$ cm$^{-2} < N < 10^{21}$ cm$^{-2}$, the gas is heated by
a combination of FUV grain photoelectric heating and X-ray photoionization heating.  It
is cooled mainly by [OI] 63 $\mu$m, [OI] 6300 \AA , [NeII] 12.8 $\mu$m, [ArII] 7 $\mu$m,
and gas-grain collisions (see Gorti \& Hollenbach 2008). The resultant temperature
is $\sim 1000-2000$ K, dropping with increasing $r$ and  $N$ as dilution and attenuation of the X-rays and FUV
lower the heating rates.  Because $T$,  $f_{Ne}(Ne^+)$, and $x_e$ 
drop with increasing $N$ (see Fig. 6) and $r$, most of the [NeII] 12.8 $\mu$m emission arises from
$r< 20$ AU and  $N\sim 10^{20} - 10^{21}$ cm$^{-2}$,  where $T\sim 1000$ K, $f(Ne^+) \sim 10^{-1} $, and $x_e \sim
 10^{-3}$.

In addition to showing the radial origin of the [NeII] emission in the EUV layer, Figure 4
also shows the [NeII] ``luminosity" plotted as a function of $r$ for the X-ray layer (hard spectrum
X-rays).
Note that there is greater contribution from inner ($\sim 1$ AU) regions of the disk compared
to the EUV layer.  In addition, there is more luminosity emerging from the X-ray layer
than the EUV layer.
Figure 4 shows that the X-ray-induced emission arises mostly from $r\sim 1-10$ AU, also
in agreement with Glassgold et al. Beyond this radius, the X-ray and FUV heating is
insufficient to maintain significant quantities of $T\gta 1000 $ K gas.

Figure 7 plots the [NeII] 12.8 $\mu$m and [NeIII] 15 $\mu$m emission from the X-ray layer.
In agreement with Glassgold et al (2007) and Meijerink et al (2008) we find that
[NeIII]/[NeII] $\lta 0.1$, mainly caused by the rapid charge exchange of atomic H with
Ne$^{++}$.  We also plot outlines of the observed 54 sources tabulated by G\"udel et al
(2009).  The vertical dotted lines shade the region where the sources have known outflows
and jets.   The horizontal solid lines shade the regions with no outflows or jets detected.
The $L_X$ tabulated by G\"udel et al is a two component fit with an attempt
to correct for absorption of the softer X-rays by material on the line of sight from star
to observer.  However, many of the observations do not extend to $h\nu < 0.3$ keV
and extinction is severe at the lower energies, so that a luminous soft X-ray source that
is weak at 0.5-1 keV could exist undetected.  The effect of such a "soft" X-ray component 
would be to move the data points to the right on Figure 7 and comparison should be
made to our "soft" X-ray spectrum results (dashed line).  
We also find that $L_{NeII}$ and $L_{NeIII}$ scale with $L_X$, 
as predicted in \S 3 and also as found by Meijerink et al (2008).  Comparison with
Figure 2 shows that if the X-ray luminosity is about the same as the EUV luminosity
from the central star, and if the EUV spectrum is soft enough that [NeII] dominates [NeIII]
in the EUV layer, the [NeII] luminosity is roughly 2 times stronger from the X-ray layer as
from the EUV layer, as we estimated analytically.  The main conclusion
from comparing the data to the model results is that although the X-ray layer may
explain the origin of the [NeII] emission in some (perhaps most if a strong soft
X-ray excess is common)  sources, there are a significant
number of sources, especially those with observed outflows and jets, where the
X-ray luminosity seems insufficient to explain the [NeII] luminosity.   
In \S 4.4 below we compare the observational data with our analytic results
on the [NeII] luminosity expected from internal shocks in the jets or winds,
and find that this is a plausible origin for these sources.

As noted above, our results on the IR fine structure emission from the
X-ray layer do not differ appreciably from 
previous results (Glassgold et al 2007,
Meijerink et al 2008).  Overall, we tend to find somewhat (factor of $\sim 2$) lower
IR line luminosities.  This agreement is a bit fortuitous, arising from a cancellation
of several effects and the insensitivity of the fine structure lines with variations
in $T$ if $T \gta 1000$ K.  Our models self-consistently calculate the vertical density
structure of the gas by using the computed gas temperatures (which differ from
the dust temperature)  to calculate the gas
density structure rather than assuming the gas density structure is fixed by
the calculation of vertical pressure balance when one assumes
the gas temperature to equal the dust temperature,  as done in the previous work.
Our self-consistent model produces significantly different results at columns 
$N \lta 10^{21}$ cm$^{-2}$, where the gas temperature
rises above the dust temperature (Gorti \& Hollenbach 2008).  The net effect is
that our gas disk is more flared, intercepting a larger fraction of the X-ray luminosity.
This tends to raise the emission from our models.  In addition,
we include FUV grain photoelectric heating which also raises the emission.
However, counteracting these effects is the inclusion of more gas coolants in
our model, especially [NeII] 12.8 $\mu$m and [ArII] 7 $\mu$m.  In addition, our treatment
of the gas heating by X-rays follows Maloney et al (1996), which is somewhat different than
the approach used by Glassgold et al and Meijerink et al, and our X-ray heating rates are
lower than these authors by a factor 3-10.  We believe this may arise
because we include the loss of "heat" due to escape of Lyman alpha and other
photons created by recombining hydrogen or to 
the absorption of these photons by dust.   Overall, our X-ray
layer tends to be a factor of about 2 cooler than the previous models  (roughly
1000-2000 K versus 2000-4000 K in the previous models),
 thereby lowering the fine structure emission from this layer.  This lower temperature,
 has a relatively small effect on the fine structure lines, because their upper states  lie
 only $\Delta E/k \sim 1000$ K above the ground state.   However, it has an enormous effect
 on our predictions of the [OI] 6300 \AA \ emission, whose upper state lies $\Delta E/k \sim
 23,000$ K above the ground state, as we will discuss below in \S 4.5.


\subsection{[NeII] Emission from Internal Shocks in the Jets and Winds}

Figure 8 plots the [NeII] luminosity versus the mass
accretion rates assembled by G\"udel et al (2009).  As in Figure 7, the vertical dotted lines shade the
region which includes sources with known jets or outflows, whereas the solid horizontal
lines denote sources with no detected jets/outflows.   We plot here our predicted [NeII] luminosities
from internal shocks in the winds/jets, using our analytic expression (Eq. 33).   The  solid line
represents the expected [NeII] luminosity when $\dot M_w = 0.1\dot M_{acc}$, the entire
wind or jet passes through a shock or  $f_{sh} = 1$, the shock velocity is in excess of about
100 km s$^{-1}$, and the preshock density is less than $10^4$ cm$^{-3}$.   The upper dashed line
makes the same assumptions except that $\dot M_w = \dot M_{acc}$ and the lower dashed line
assumes $\dot M_w = 0.01\dot M_{acc}$.   Note that the [NeII] luminosity is proportional to
the product of $f_{sh}$ and $\dot M_w$, so that, for example, the lower dashed line
also corresponds to $f_{sh} = 0.1$ and $\dot M_w = 0.1\dot M_{acc}$.    The main conclusion
is that internal wind or jet shocks very likely explain the origin of the [NeII] from the outflow
and jet sources.   In fact, the figure might suggest that these shocks could explain the [NeII] observed
in nearly all of the sources, if it were not for the fact that in some cases (e.g., Herczeg et al 2007, Najita et al 2009, Pascucci et al 2009) where the lines have been spectrally resolved, they are narrower than a shock origin would predict.    We do note that in some of these
cases, the integrated flux seen with the high spectral and spatial resolution ground based instruments is
significantly less than the flux seen by the low resolution {\it Spitzer Space Telescope}.
Najita et al speculate that perhaps there are two components comprising the total flux, a strong
but broad and extended shock component and a weaker, but narrow and spatially unresolved
disk component arising from the X-ray layer.
 On the other hand, it is quite possible that
X-rays or EUV produce most of the [NeII] luminosity in the sources with no observed winds or jets.
Note that these sources in Figure 8 are distributed in a nearly horizonal line with no apparent
dependence on $\dot M_{acc}$ over two orders of magnitude increase in this parameter.

\subsection{[OI] 6300 \AA \ Emission from the EUV and X-ray Layer}

Figure 9 plots the [OI] 6300 \AA \ luminosity from the EUV layer 
versus $\Phi _{EUV}$ for both our
harder $L_{EUV}(\nu) \propto \nu ^{-1}$ spectrum and our  softer $L_{EUV}(\nu)$ 
blackbody spectrum.   A harder spectrum gives more [OI] luminosity in the EUV
layer because, although the gas is almost entirely ionized, there is a greater
fraction of neutral H and O in the gas due to the smaller photoionization
cross section of these atoms with higher photon frequency.   However, even the harder
EUV spectrum results in [OI] 6300 \AA \ luminosities $\la 10^{-6}$ L$_\odot$, which can
only explain the weakest LVC sources.  Recall that $L_{[OI]}$ ranges from $10^{-6} -
10^{-3}$ L$_\odot$ in LVCs.

Figure 10 plots the  [OI] 6300 \AA \ luminosity  in the X-ray layer versus $L_X$ for both
our harder and our softer X-ray spectrum.   The [OI] luminosity
does increase with $L_X$, due to the higher temperatures and higher ionization
fractions in the X-ray layer.   Recall that the gas is primarily neutral, so the
higher ionization fraction increases the luminosity by increasing the density
of the electrons which excite the [OI].  However, with our standard (harder) X-ray spectrum,
which is quite similar to that adopted by Meijerink et al (2008), we obtain [OI] 
luminosities that are a factor of nearly 100 times lower than those
of Meijerink et al (2008) from the X-ray layer.  This is primarily because of the extreme
sensitivity of the [OI] luminosity to the temperature of the X-ray layer (see Eq. 38).  As noted
above, our temperatures are roughly a factor of 2 lower than those of Meijerink et al.

The main point, however, is that in our standard models neither the EUV layer or the X-ray layer
can produce [OI] luminosities as high as $10^{-5} - 10^{-3}$ L$_\odot$ as
observed in many LVC sources.    This confirms the analytic estimates made
in Section 3.5.   However, as discussed above, the [OI] luminosity is very sensitive
to the temperature of the X-ray layer.   One way to increase the temperature is
to assume a softer X-ray spectrum.   Softer X-rays have much higher absorption
cross sections, and therefore deposit much more heat per unit volume in the
upper layers.   In addition, they create higher electron abundances, and these
lead to increased efficiency in converting the absorbed X-ray energy into heat.
Therefore, we also plot in Figure 10 the results for cases with similar X-ray luminosities,
but with our "soft" X-ray spectrum where  $L_\nu \propto \nu ^{-1}$ from 0.1 keV
to 2 keV.   This spectrum has
many more 0.1-0.3 keV X-rays than our standard case, and we find that we do indeed
get higher temperatures and electron abundances in the upper parts of the X-ray layer,
and consequently much higher [OI] luminosities.  $L_{[OI]}$ can be as high as $\sim 10^{-4}$
L$_\odot$ if $L_X \sim 10^{32}$ erg s$^{-1}$ ($\sim 2\times 10^{-2}$ L$_\odot$), a likely upper limit to the 
soft X-ray luminosity.
Therefore, soft X-rays may be able to explain "typical", $L_{[OI]} \sim 10^{-4}$ L$_\odot$, LVC
sources, but not the most luminous sources.
We note that if the soft X-rays were causing photoevaporation, then high spectral resolution
observations of 
the [OI] line might diagnose the flow parameters.

\section{DISCUSSION}

This paper has focused on fine structure lines from ions which required
$h\nu > 13.6$ eV photons to photoionize them.  Most of these lines fall in
the 5 $ \mu$m $< \lambda <$ 40 $\mu$m wavelength region and are therefore
partially accessible through atmospheric windows from the ground, and entirely
accessible from space-based observatories such as the {\it Spitzer Space
Telescope}.  Ground based observatories
have the advantage of larger diameter telescopes and therefore greater spatial resolution
as well as larger and heavier instruments capable of higher spectral resolution.
The TEXES instrument (Lacy et al. 2002) achieves a spectral resolution of $\sim 3$ km s$^{-1}$ and a spatial resolution of $\sim 0.5 (\lambda /10 \mu m)$ arc seconds on a 10 meter-class ground-based  telescope such as Gemini.
Its sensitivity to line flux (5 sigma in one hour) translates to $L \sim 3\times 10^{-7}$ L$_\odot$ at 100 pc.   The Michelle instrument (Glasse et al. 1997)  achieves spectral resolution of $\sim 15$  km s$^{-1}$  and is capable of detecting lines with luminosities
$L \sim 3\times 10^{-6} $  L$_\odot$
at 100 pc, if mounted on a 10 meter-class ground-based telescope.  The VISIR instrument on an 8 meter class
telescope has a sensitivity of $\sim 3\times 10^{-6}$ L$_\odot$ at 100 pc, and a spectral resolution of about
12 km s$^{-1}$ (Lagage et al 2004).   {\it Spitzer} had relatively poor spatial ($\sim
12$ arcsec resolution ) and spectral  ($\sim 500$ km s$^{-1}$) resolution but could achieve 5 sigma in one hour sensitivity that translated to $L \sim 10^{-7}$ L$_\odot$ at 100 pc.  

A number of groups have now observed nearby star-disk systems and measured fluxes from
especially the [NeII] 12.8 $\mu$m line, with a few detections of H I reco mbination lines 
and good upper limits for [NeIII] 15 $\mu$m lines.  Many of the observations were done using the IRS
spectrometer on {\it Spitzer} (Espaillat et al 2007, Lahuis et al 2007, Pascucci et al 2007, Ratzka et al 2007). 

However, we first discuss recent ground-based observations with high resolution spectroscopy that help constrain
 the origin of the NeII emission by interpretation of the observed linewidths and spatial extents, as well as
 by the observed fluxes (e.g., Herczeg et al. 2007, van Boekel et el. 2009, Najita et al. 2009, Pascucci \& Sterzik 2009). The first such resolved source to be observed and also one of the brightest  is TW Hya  (Herczeg et al. 2007).  
Herczeg et al interpret the observed line width ($\sim 21$ km s$^{-1}$) from this nearly face-on
disk as possibly indicating the emission arises from the inner regions ($\sim 0.1$ AU)
of the disk.  In our models it is very difficult to produce the observed [NeII] luminosity from
X-rays or EUV at 0.1 AU.  However, as they note, it might also originate from the EUV or X-ray layers
at $r \sim 10$ AU, if turbulence can produce the observed linewidths.  Alternatively, the linewidth 
may arise from the fact that the gas is not
merely in Keplerian rotation, but also is photoevaporating at $\sim10$ km s$^{-1}$ with
respect to the disk surface.  This produces a blue shift of the [NeII] with respect to the
stellar velocity (e.g., Alexander 2008).  Using the VISIR spectrograph on the VLT telescope {\it Melipan}, Pascucci \& Sterzik (2009) recently showed that nearly all the line flux is blueshifted, with a peak at -6.3 km s$^{-1}$ 
and  a FWHM of 14.2 km s$^{-1}$.  They point out that these observations are in near perfect
agreement with the prediction of Alexander (2008) for  [NeII] produced in EUV-induced photoevaporating
flow, and inconsistent with a static disk atmosphere.  Alternatively, a soft X-ray spectrum might produce a very
similar photoevaporating profile, since soft X-rays heat the disk surface at 1-10 AU  to almost the same temperatures
as the EUV layer.   We note that TWHya is known to have a strong soft X-ray excess (Kastner et al 2002). 
The measured low 
accretion rate, $5\times 10^{-10}$ M$_{\odot}$ yr$^{-1}$ (Muzerolle et al. 2000), and the absence of any known outflow support an EUV and/or X-ray heated disk origin for the [NeII] emission.  The observed [NeII] luminosity as measured
by Pascucci \& Sterzik is $\sim 4\times 10^{-6}$ L$_\odot$.
From our models, we predict no contribution from shocks which is consistent with the low observed linewidths. We would expect EUV and X-rays to irradiate the disk given the low accretion rate and the measured flux to be a sum of the contributions from the ionized and neutral layers of the disk. We calculate the contribution from the neutral layer to be $\sim 3 \times 10^{-6}$ L$_{\odot}$  (using $L_X = 2\times 10^{30}$ erg s$^{-1}$; Kastner et al. 2002). If the remaining $\sim 10^{-6}$ L$_{\odot}$ is from the ionized layer,  we then estimate that $\Phi_{EUV}=3 \times 10^{40}$ s$^{-1}$ for TW Hya. However,  given the accuracy of our models, we cannot rule out that most of the [NeII] emission is from
the EUV layer ($\Phi_{EUV}=1.2 \times 10^{41}$ s$^{-1}$) .   The excellent agreement of the [NeII] line profile with
the EUV model suggests that EUV may dominate in this source, but our model results using the observed Xray
luminosity suggests that a substantial amount of the [NeII] may arise in the X-ray layer.    Modeling of X-ray induced flows and further observations are
needed to clarify this discrepancy, possibly of the [ArII] 7 $\mu$m line which might discriminate between 
[NeII] emission from EUV or X-ray layers (see discussion below at end of this section).

 Herczeg et al. fail to detect [NeII] emission for the sources BP Tau and DP Tau.  These non-detections are also compatible with the [NeII] emission models described in this paper. BP Tau is an actively accreting star ( $\sim 2\times 10^{-8}$ M$_{\odot}$ yr$^{-1}$; Muzerolle et al. 2000) with presumably no EUV penetration of the disk wind and an X-ray luminosity ($L_X \sim 7\times 10^{29}$ erg s$^{-1}$) that would produce lower [NeII] emission  than the upper limit from observations.  DP Tau has a low accretion rate, but very poor upper limits to the [NeII] flux to provide any reasonable estimates of $\Phi_{EUV}$. 

Van Boekel et al. (2009) report that the [NeII] emission from the T Tau triplet, which is resolved spatially and spectrally, has large linewidths $\sim 100$ km s$^{-1}$ and is associated with a known outflow. Even though T Tau N is a very strong X-ray source with   $L_X\sim 2 \times10^{31}$ erg s$^{-1}$ (G\"{u}del et al. 2007), the expected [NeII] from the disk is still a factor of $\sim 10$ lower than what is observed. In addition, van Boekel et al spatially resolve the [NeII] emission and
determine that it arises from T Tau S.
We do not expect EUV and soft X-rays to penetrate the disk wind for this young source.  On the other hand, our
models of shock emission are consistent with the van Boekel et al data, as  discussed in \S 3.4.   van
Boekel et al  also conclude that shock emission is the likely origin of the [NeII] emission.

Najita et al. (2009) have observed two young disks around AA Tau and GM Aur using TEXES on {\it Gemini N}, and spectrally resolved the [NeII] line in both the sources.
 The FWHM linewidths are 70 and 14 km s$^{-1}$, respectively, and the authors interpret the emission as arising from the X-ray layer in Keplerian disks.  They also note that the flux in the line is less than that measured by the much larger beam of {\it Spitzer}.  A spatially extended and broad (FWHM) additional component, such as a protostellar wind shock, could account for the difference. GM Aur is a transition disk object, that is still actively accreting at $\sim10^{-8}$M$_{\odot} $yr$^{-1}$, indicating the presence of gas in the dust depleted inner disk. The disk accretion rate is at a marginal epoch where the EUV may make it through to irradiate the disk or may be absorbed by the disk wind.  From the known X-ray luminosity of the star ($L_X\sim10^{30}$erg s$^{-1}$; Strom et al. 1990), we estimate an X-ray produced [NeII] line luminosity of $2\times 10^{-6}$ L$_{\odot}$, while the observed value is  $\sim 7\times 10^{-6}$ L$_{\odot}$ (Najita et al. 2009).  The rest may arise from shocks, although no known outflows exist. Alternately it may come from either
 an unobserved EUV or soft X-ray component that has just begun to penetrate the disk wind and heat
 and ionize the surface layers.  Note that if EUV dominates , $\Phi_{EUV} \sim 2\times 10^{41}$ s$^{-1}$.  The classical T Tauri  star, AA Tau has a low accretion rate for an  object of its class, estimated at $3\times 10^{-9} $M$_{\odot}$ yr$^{-1}$ (Gullbring et al. 1998), and we expect irradiation of the disk by EUV and X-ray photons due to the expected low wind column density. The observed line luminosity is  $\sim 4\times10^{-6}$ L$_{\odot}$. AA Tau is highly X-ray variable with $L_X\sim 3\times10^{30} - 2 \times 10^{31}$ erg s$^{-1}$ (Schmitt \& Robrade 2007) which can result in NeII luminosities arising from the X-ray heated neutral layer, ranging from $ 10^{-6}$L$_{\odot}$ to $10^{-5}$L$_{\odot}$, and the observed value lies within this range.  While it is likely that the X-ray layer explains the origin of the [NeII], the observed
 [NeII] flux places an upper limit of $\Phi_{EUV} \lta 2 \times 10^{41}$ s$^{-1}$ for AA Tau.

Pascucci \& Sterzik (2009) detect [NeII] in all the three transition disks that they observe (TW Hya, CS Cha, T Cha), but only from one of the three classical disks (Sz 73). They claim that the resolved linewidths of all the transition disks are consistent with a photoevaporative flow driven by stellar EUV photons and estimate $\Phi_{EUV} \sim 10^{41-42}$ s$^{-1}$.  These numbers should be considered as upper limits to $\Phi_{EUV}$ since there may be some
contribution to the [NeII] flux from the X-ray layer.   Pascucci \& Sterzik also observe blueshifted [NeII] emission in CS Cha and T Cha, consistent with EUV photoevaporation.  In CS Cha the inferred hole size is 45 AU.  We note that if the
inner disk is completely clear of gas, such a large hole is only consistent with EUV-induced [NeII] emission since
the X-ray flux at this radius is too low to heat the gas to temperatures $\gta 1000$ K required to excite the [NeII].  
Pascucci \& Sterzik point out that they only detect such evidence of EUV photoevaporation in sources with very low accretion rates, consistent with  our model here that EUV does not penetrate the wind base until the accretion rates are low. We also note that the expected X-ray heated [NeII] emissons for their non-detections are consistent with their upper limits.

We next discuss the totality of [NeII] observations, which is dominated by unresolved {\it Spitzer} sources.
There is clearly a considerable amount 
of scatter when one tries to see if the [NeII]  luminosities $L_{NeII}$ correlate with either
the X-ray luminosity $L_X$ or with  $\dot M_{acc}$.   Espaillat et al. (2007)  conclude that the [NeII] has a nearly  linear correlation  with the mass accretion rate; they find a 10 times increase in [NeII] luminosity with about
a 10-fold increase in accretion rate.\footnote{We note that Pascucci et al (2007) 
found a tentative anticorrelation with accretion rate, but this was based on a very
limited dataset which had a small range in values of line luminosity and
accretion rates.} However, Espaillat et al had a data set of only 7 sources, whereas
recently, G\"udel et al (2009) have compiled a data set of more than 50.
 G\"udel et al find little $L_{NeII}$ dependence
with mass accretion rates at low  $\dot M_{acc} \lta 3\times 10^{-8}$ M$_\odot$ yr$^{-1}$,
but a roughly linear trend in the high  $\dot M_{acc} \gta 3\times 10^{-8}$ M$_\odot$ yr$^{-1}$
sources which show evidence for jets and outflows (see Figures 7 and 8).   The latter suggests that protostellar
wind shocks may be responsible for the [NeII] from the outflow sources.   It seems unlikely that the [NeII]
in these sources is due to soft X-rays or EUV, since the wind mass loss rates are
sufficiently high to likely block these photons from ever striking the disk surface at
radii near $r_g$.  In addition, our analytic predictions of [NeII] luminosities from wind
shocks seem to match the observations (Figure 8).
Espaillat et al find little correlation of $L_{NeII}$  with $L_X$.  G\"udel et al
formally find $L_{NeII} \propto L_X^{0.58}$ but with a tremendous amount of scatter.
We note that although many of the observed X-ray luminosities derive from observations of $\sim 0.2 
-10$ keV
X-rays, the soft (0.1-0.3 keV) X-rays may suffer considerable extinction that is difficult to estimate,
and considerable luminosity could be "hidden" in such a soft component.
 Some of the observed scatter may then
be caused by  [NeII] arising from EUV, soft X-ray, or shock heated and ionized gas.

In summary, shocks may dominate at high  $\dot M_{acc} \gta 3\times 10^{-8}$ M$_\odot$ yr$^{-1}$,
but there is  observational evidence that EUV or X-rays must dominate at
lower accretion rates.  Because X-rays are more efficient in producing [NeII], naturally produce
[NeII] stronger than [NeIII] as observed, and more easily penetrate the base of the
protostellar wind, it seems likely that X-rays often dominate the EUV production
of [NeII] in disks, although not by a large factor.
A part of this evidence for a non-shock origin has been gathered by high spectral
resolution observations of [NeII] made by ground-based telescopes, which show
relatively small linewidths compared to the $\gta 100$ km s$^{-1}$ linewidths expected
for wind-shocked [NeII].  Although $\sim 1$ keV X-rays may play a role in the
production of $L_{NeII}$ for sources with weak winds, there is clear evidence that EUV or soft X-rays
may sometime dominate.
If one wanted to identify a source where it is likely that either EUV or soft X-rays
dominate the [NeII] production, one would choose sources with low accretion rates,
 $ \dot M_{acc} \lta 8 \times 10^{-9}$  M$_\odot$ yr$^{-1}$, whose $L_{NeII}$ lies well
 above the observed correlation of $L_{NeII}$ with the 1 keV $L_X$.


One of our principal results is that X-rays are more efficient in producing [NeII] emission
than are EUV photons.    If the central star has the same luminosity in X-rays
as it does in EUV photons, the [NeII] luminosity from the X-ray layer will be about 2 times
greater than the [NeII] from the EUV layer (assuming a soft EUV spectrum, which is
the most efficient in producing [NeII]).   This result was shown both analytically,
in Section 3, and in our numerical results as seen in Figures 2 and 7.  Since the
[NeII] luminosity scales linearly with the EUV luminosity and with the X-ray luminosity,
this means that the EUV luminosity needs to be at least 2 times the X-ray luminosity
for the EUV to dominate the production of [NeII].   Unfortunately, we have little idea of
the EUV luminosity, since it is impossible to observe in young sources.
Observations of older, nearby stars by Ribas et al (2005) suggest that
the luminosity in the EUV band is usually similar to that of the X-ray band.   However,
these sources are not accreting, and it is possible for an accreting source to
be very bright in EUV relative to 1 keV X-rays (but see Alexander 2004b and Glassgold
et al 2009, who argue these photons are attenuated by the accretion columns near the star).   
In any event, these accreting
stars need to have sufficiently low wind mass loss rates to allow these
accretion shock-generated EUV photons to penetrate the wind base and strike the
outer disk to create EUV-generated [NeII].   Alexander et al (2005) estimate
EUV fluxes from stars with observed ultraviolet emission lines, and conclude 
that in some cases the EUV photon luminosities can be as high as $10^{44}$ s$^{-1}$.
This suggests that in some cases, the chromospheric emission may generate more
EUV luminosity than X-ray luminosity in young stars.  However, our own results
place upper limits on the possible EUV photon luminosities:
$\Phi _{EUV} \lta 10^{42}$ s$^{-1}$.  Overall, it appears that it is unlikely that the EUV 
fluxes on the disk surface are any stronger than the X-ray fluxes, and that it is
likely that X-rays often slightly dominate EUV photons in the production of [NeII] when the wind mass loss rates
are low so that internal wind shocks are weak.

We have plotted the observed [NeII] and [NeIII] data on Figure 7, using the compilation of G\"udel et al (2009)
which uniformly treats all previously observed sources.  The observed
[NeIII]/[NeII] ratio of less than 0.06 in the source with measurements of
both lines (Sz102, Lahuis et al. 2007)  favors either an origin in the X-ray layer, a shock, 
or in a soft ($T_{eff} \lta 40,000$ K) EUV layer.  A hard EUV layer such as
our adopted power law  $F_\nu \propto \nu ^{-1}$ is ruled out.  Several of the sources
have [NeII] luminosities readily explained as arising in the X-ray regions, as noted by
Meijerink et al (2008).  However, a number of the sources have larger [NeII] luminosities
than can be explained by $\gta 0.5$ keV X-rays alone.  In many such cases, such as the T Tau South source
discussed by van Boekel et al (2009), shocks in the protostellar wind are the likely source.
We note that since wind mass loss rates scale with accretion rates, shocks would provide
the observed correlation (Espaillat et al 2007, G\"udel et al 2009) between the mass accretion rate and
$L_{[NeII]}$ (see Figure 8). 

If the wind mass loss rates are not sufficient to provide the observed
[NeII] luminosity, or if ground-based observations reveal narrower lines than might
be expected from the shocks, such as in TW Hya, [NeII] emission may be generated by
a soft EUV  or X-ray spectrum from the central star.  
Since the "hard" X-rays were insufficient to explain some of these sources, and since we have
shown that X-rays are more efficient in producing the [NeII] line, the only way
that EUV luminosity from the central star can explain these sources is for the EUV
luminosity to be greater ($>2$ times) than the observed X-ray luminosity, and,
in addition, the EUV spectrum has to be "soft" ($T_{eff} \lta 40,000$ K).  If EUV does dominate, 
we can see from Figures 1 and 7 that $L_{EUV} \lta 10^{-2}$ L$_\odot$  is often required.
 A luminosity of $10^{-2}$ L$_\odot$  corresponds to $\Phi _{EUV} \sim 10^{42}$ s$^{-1}$. 
 The comparison of the [NeII] and [NeIII] data
with Figures 1 and 2 gives hard upper limits on $\Phi _{EUV}$.  Most sources have
$\Phi _{EUV} \lta  10^{42}$ s$^{-1}$.  If EUV is the main excitation
mechanism, the comparison actually measures $\Phi _{EUV}$ and the [NeII]/[NeIII] ratio
constrains the EUV spectrum. 

We have examined our model results for diagnostics that would reveal whether the
[NeII] emission arises from the EUV layer or from the X-ray layer.  One possible diagnostic
is the ratio of the [NeII] 12.8 $\mu$m line to the [ArII] 7 $\mu$m line, [NeII]/[ArII].   We have
shown in Figures 2 and 3, along with the analytic calculation Eq. (8) combined with Table 1,
that in the EUV layer  the [NeII]/[ArII] ratio is about unity for our "soft" EUV spectrum.   This
spectrum
maintains most Ne and Ar in singly ionized form in the EUV layer.  Although elemental Ar is 20 times
less abundant in the HII gas than Ne, the rate coefficient for electronic excitation of the
[ArII] line is about 10 times larger than that of [NeII], and the 7 $\mu$m line has almost
twice the photon energy as the 12.8 $\mu$m line, making up for the abundance discrepancy.
In the X-ray layer, most of the Ne and Ar is neutral, and the fractional abundance of Ne$^+$
and Ar$^{+}$ depends, in addition to elemental abundances, on the X-ray photoionization 
cross sections of Ar and Ne, on
the electron rate coefficients for collisional ionization of Ar and Ne by secondary electrons,
and on the rate coefficients for electronic recombination of Ne$^+$ and Ar$^+$.  In addition,
the [ArII] line lies $\Delta E/k \simeq 2060$ K above ground, whereas the [NeII] line lies
only $\simeq 1100$ K above ground.   Since the X-ray heated gas is typically $\sim 1000$ K,
this means that the relative line strengths are sensitive to the temperature of the X-ray heated
layer, with [ArII] gaining advantage in warmer gas relative to [NeII].   We find in our models
that for our hard X-ray spectrum, which peaks at 2 keV and where the Ne and Ar are ionized
mainly by direct X-ray photoionization, the X-ray layer produces [NeII]/[ArII]$\simeq 2.5$.
Unfortunately, due to a coincidence of atomic parameters and the enhanced heating
due to soft X-rays,  for our soft X-ray spectrum
the ratio is [NeII]/[ArII]$\simeq 1$, {\it the same as in the EUV layer}.   Thus, this ratio
may discriminate between [NeII] produced in the X-ray layer versus the EUV layer only
when the X-ray spectrum is relatively "hard".  Nevertheless, a large ratio would strongly
point to an origin in the X-ray layer.

We have also examined both analytically and numerically the expected [OI] 6300 \AA \ 
luminosity from disks around young stars.  The observed luminosities in this line range
from 10$^{-6}$ to 10$^{-3}$ L$_\odot$ in the low velocity component, which has been identified
as arising from the disk.  We have shown that the EUV, transition, and (hard) X-ray layers are
not likely to produce [OI] 6300 \AA \  luminosities greater than 10$^{-6}$ L$_\odot$.
Meijerink et al (2008) provided models utilizing a relatively "hard" X-ray spectrum (peaking
around 1 keV) which achieved [OI] luminosities as high as
$\sim 10^{-4}$ L$_\odot$, but our models with a similar X-ray spectrum give [OI]
luminosities $\sim 10^{-6}$ L$_\odot$.    The [OI]  6300 \AA \ line is extremely sensitive to the
temperature in the X-ray layer, as we showed analytically in Section 3.5.   Our models give
typical temperatures of 1000 - 2000 K, whereas the Meijerink et al models give 2000-4000 K.
We discussed in \S 4.3 the improvements in our models which lead to lower gas
temperatures in the X-ray layer.  However, Ercolano et al (2009) appeal to observational constraints on
the emission measure distribution as a function of temperature for the chromospheres
of young star analogs to argue that there is a (largely unobserved) soft X-ray component
that is much larger than that assumed in our standard X-ray spectrum and in Meijerink
et al (2008).
Ercolano et al find that the X-ray spectrum may be better approximated by a power law $L_\nu
\propto \nu ^{-1}$ from 0.1 keV to 2 keV.  We have also run cases with such a soft
X-ray spectrum, and find that X-ray luminosities of $\sim 10^{32}$ erg s$^{-1}$
can then give rise to [OI] luminosities of $\sim 10^{-4}$ L$_\odot$.  

\section{SUMMARY AND CONCLUSIONS}

Circumstellar disks around low mass stars evolve with time with a decreasing accretion rate
onto the star and a decreasing wind mass loss rate from the inner disk.   X-rays, EUV and FUV
photons from young, low mass stars arise principally from either magnetic activity (an active
chromosphere) or from the accretion shock arising as disk material falls onto the star, presumably
in accretion columns along stellar magnetic field lines.  In the latter case, the energetic photons must
penetrate or obliquely avoid the accretion columns in order to illuminate the disk surface.
In either case, they must penetrate the protostellar wind near the wind base.  We treat here the
penetration of the protostellar wind and find that FUV photons likely penetrate first, when the wind
mass loss rate is $\dot M_w \gta 4 \times 10^{-8}$ M$_\odot$ yr$^{-1}$, the exact number depending
on the very uncertain dust opacity in the wind base material.  As the wind mass loss rate drops with
time, $\sim 1$ keV X-rays penetrate next, when $\dot M_w \simeq 4 \times 10^{-8}$ M$_\odot$ yr$^{-1}$.
Finally, soft ($\sim 0.1$ keV)  X-rays and EUV photons penetrate only when the wind can be fully ionized
at the base, which occurs roughly at $\dot M_w \lta 8 \times 10^{-10}$ M$_\odot$ yr$^{-1}$.
The corresponding mass accretion rates onto the star are about 10 times higher, with considerable scatter.
Considering observed rates of mass accretion with time (e.g., Hartmann et al 1998), these criteria
translate to FUV and 1 keV X-rays penetrating very quickly after mass infall onto the disk from
the molecular core has ceased, whereas EUV and soft X-rays may require an additional 1-2 Myr (with a lot
of scatter) before they illuminate the disk.

The 1 keV Xrays and FUV photons penetrate the disk surface to vertical columns of $N\sim 10^{21}$ cm$^{-2}$,
and heat this layer to temperatures of order 1000 K for $r\lta 10- 20$ AU.  The X-rays ionize hydrogen
and atoms with ionization potentials (IP) $> 13.6$ eV in this predominantly neutral layer, providing both
electrons and species such as Ne$^+$ and Ar$^+$.  Thermal collisions of the electrons with these
species produce fine structure lines such as [NeII] 12.8 $\mu$m.  The high gas temperatures and
elevated electron abundances also produce strong emission from the [OI] 6300 \AA \ forbidden line
in regions with $T \gta 2000$ K.  The FUV photodissociates molecules,  ionizes species with
IP $<13.6$ eV, and contributes to the gas heating.

The EUV photons incident upon the disk create a fully ionized (HII) layer with $T\sim 10^4$ K, that lies
above the X-ray layer on the disk surface.   Here,
EUV photoionizes species with IP $\gta 13.6$ eV, and singly or doubly ionized species
tend to be the dominant ionization stage.  Trace amounts of atomic oxygen are present and a relatively 
small amount of [OI] 6300 \AA \  luminosity emerges from this layer.  Due to a combination of falling
electron density, rising scale height, and increasing disk surface area with increasing $r$, most
of the fine structure emission from the EUV layer arises from $r\sim r_g \sim 7$(M$_*$/1 M$_\odot$) AU. 
The EUV layer produces more hydrogen recombination line luminosity than the X-ray layer, but
does not explain the observed high ratio of these lines to [NeII].   It is likely the hydrogen recombination
lines are produced in dense plasma close to the star: in the chromosphere, the accretion shock, or in a wind shock very
close to the star.

Strong ($\gta 100$ km s$^{-1}$) shocks, such as can be produced in internal shocks in protostellar
winds or jets, can also significantly ionize species with IP $> 13.6$ eV and heat the gas to $T>>1000$ K,
sufficient to excite the fine structure lines, the hydrogen recombination lines, and optical forbidden lines
such as [OI] 6300 \AA .  Such ionization and heating has been inferred by observation of optical lines emitted
in knots in the jets and in Herbig-Haro objects.

In this paper we have analytically modeled all three of these emitting regions, and have presented results
from detailed thermo/chemical numerical models of the EUV and X-ray layer.  We have focussed on the
emergent line luminosities of [NeII] 12.8 $\mu$m, [NeIII] 15.5 $\mu$m, [ArII] 7 $\mu$m, [ArIII] 9 $\mu$m, 
[SIII] 19 $\mu$m, [SIII] 33 $\mu$m, and [OI] 6300 \AA .   However, we also discussed infrared hydrogen recombination
lines (6-5 and 7-6) and other fine structure lines such as [SIV], [NII], [NIII], and [OIII].  These  line luminosities
are diagnostic of key parameters such as the EUV luminosity and spectral shape, the
X-ray luminosity and spectral shape, and the wind mass loss rate and shock speed.  Our main results
are as follows:

1. The luminosity of fine structure lines (e.g., [NeII] and [ArII])  from the dominant ionization state of a species 
roughly scale with $L_X$ and $L_{EUV}$.  At very high $L_X$ or $L_{EUV}$ the lines saturate because
the electron density in the emitting region exceeds the critical density of the line.   [ArII] 7.0 $\mu$m, which
has not yet been observed, is predicted to be as strong as [NeII] 12.8 $\mu$m in the EUV layer.  If the
X-ray layer dominates and the X-ray spectrum is such that much of the X-ray luminosity 
is in the 1-3 keV band, the [ArII] line
is predicted to be about 2.5 times weaker than the [NeII] line.  Therefore, the observed [NeII]/[ArII] flux ratio
may help determine the origin of these lines.
Observations of [NeII] set upper limits for the EUV luminosity of the central star, $\Phi _{EUV}\lta 10^{42}$
EUV photons s$^{-1}$ for most sources.

2. Most of the fine structure emission in the EUV layer arises from 5-10 ($M_*/$M$_\odot$) AU.   Most of
the fine structure emission from the X-ray layer is distributed more broadly in $r$ from $\lta 1 - 10$ AU
for a solar mass star.

3. If $L_X \sim L_{EUV}$, there is about 2 times as much [NeII] emission arising from the X-ray layer
as from the EUV layer, assuming our standard "soft" EUV (30,000 K blackbody)
spectrum which produces the most [NeII] luminosity.

4. A power-law EUV spectrum, $L_\nu \propto  \nu ^{-1}$, results in a [NeIII] line luminosity that
is greater than the [NeII] line luminosity from the EUV layer, contrary to observations.  If the EUV layer
is responsible for the [NeII] emission, the EUV spectrum must be softer than a $\sim 30,000 - 40,000$ K
blackbody spectrum between 15 eV and 40 eV.  The X-ray layer, which has much higher abundances
of atomic hydrogen, naturally gives [NeIII] line luminosities that are less than 0.1 of the [NeII] luminosities 
because of rapid charge exchange reactions of Ne$^{++}$ with H.

5. Internal shocks in protostellar winds may be a viable explanation of the observed [NeII] in a number
of sources, especially those with high $\dot M_w$ or its surrogate $\dot M_{acc}$.  Confirmation of this
origin requires high spatial ($\lta 1"$) and spectral ($\lta 10$ km s$^{-1}$) observations.  The [NeII] from
these regions, if they are nearby, may be extended ($\gta 1"$) and should produce broader ($\sim 100$
km s$^{-1}$ FWHM) profiles than the [NeII] from the EUV or X-ray layer, especially in face-on disks.

6. [OI] 6300 \AA \ is weak ($L_{[OI]} \lta 10^{-6}$) from the EUV layer, the transition layer between
the EUV layer and the X-ray layer, the X-ray layer if the spectrum is dominated by 1-2 keV photons, and
likely also from the shear layer where the protostellar wind impacts the disk surface.  A soft X-ray spectrum
($L_\nu \propto \nu ^{-1}$ for 0.1 keV $< h\nu < $ 2 keV) with considerable luminosity in 0.1 - 0.3 keV photons
produces a hotter and more ionized X-ray layer, and substantially more [OI] 6300 \AA \ luminosity
because of the extreme temperature sensitivity of this line.   $L_X$ as high as $10^{-2}$ L$_\odot$ with
this spectrum results in $L_{[OI]} \sim 10^{-4}$ L$_\odot$.  The observed values of the low velocity component
of [OI] range from $10^{-6}$ to $10^{-3}$ L$_\odot$, with typical values $\sim 10^{-4}$ L$_\odot$.  Therefore,
soft X-rays are a plausible origin for the low velocity [OI] component in many sources.

7. We compared our models with a compilation of 54 sources of [NeII] emission
from young low mass protostellar sources and with correlations of $L_{[NeII]}$ with $L_X$ and
$\dot M_{acc}$.   We note in point 5 that internal shocks in winds may be a viable explanation for
especially the sources with observed outflows or jets.   There are also sources with low $\dot M_{acc}$
where our "harder" X-ray spectrum, with most luminosity emerging at 1-2 keV, can explain the observed
[NeII] emission.   In some cases, the lines are resolved to be relatively narrow (10-60 km s$^{-1}$), further
indicating an X-ray layer origin and not a shock origin.   However, there exist sources where neither wind
shocks nor 1-2 keV X-rays carry sufficient energy to power the observed [NeII] line.  These sources are likely
candidates for [NeII] originating from the EUV layer or from an excess of soft ($\sim 0.1-0.3$ keV) X-rays.  If the spectrum
in the EUV-soft X-ray wavelength region is a power law $L_\nu \propto \nu^{-1}$, as Ercolano et al (2009) suggest,
then the soft X-ray layer will dominate the production of [NeII], although the EUV layer may produce more
[NeIII] than the X-ray layer.   Whichever layer dominates, the [NeII] and [NeIII] luminosities directly
provide a measure of the heretofore unobserved EUV or soft X-ray luminosities from the protostar or
its immediate environs.

\section{ACKNOWLEDGMENTS}
We thank R. Alexander, C. Clarke, J. Drake, B. Ercolano, A. Glassgold,  M. G\"udel, M. Kaufman, R. Meijerink,  
J. Najita, D. Neufeld and I. Pascucci for helpful discussions
and allowing us access to prepublication drafts of papers.  We also thank R. Alexander for his
helpful and thorough referee report, and E. Feigelson, the editor, for helpful comments on
X-ray flare size, time variability, and spectral shape.
We acknowledge financial support from NASA's Origins Program, Astrobiology Program,
and Astrophysical Theory Program.

\begin{deluxetable}{lcccl}
\tablewidth{4in}
\tablecaption{IR Fine Structure Parameters for Species in Ionized Gas}
\tablehead{\colhead{Transition} &\colhead{$C_t$ } &\colhead{$n_{ecr.s}$ } &\colhead{$x(s)$} & \colhead{Refs.} \\
&({\rm ergs})&({\rm cm}$^{-3}$)& &}
\startdata
$[$ArII$]$ 7 $\mu$m&1.8(-13)&4.2(5)&6.3(-6)\tablenotemark{a}  &1\\
$[$ArIII$]$ 9 $\mu$m&3.7(-14)&1.2(6)&6.3(-6)\tablenotemark{a} &2 \\
$[$NII$]$ 122 $\mu$m&2.3(-14)&1.6(3)&9.1(-5)\tablenotemark{a} &3\\
$[$NIII$]$ 58 $\mu$m&5.0(-13)&1.2(3)&9.1(-5)\tablenotemark{a} &4\\
$[$NeII$]$ 12.8 $\mu$m&2.2(-13)&6.3(5)&1.2(-4)\tablenotemark{b}  & 5 \\
$[$NeIII$]$ 15.0 $\mu$m&3.7(-13)&2.7(5)&1.2(-4)\tablenotemark{b}  & 6\\
$[$OIII$]$ 52 $\mu$m&8.3(-13)&4.6(3)&3.2(-4)\tablenotemark{a} & 3\\
$[$SIII$]$ 18 $\mu$m&3.3(-13)&1.5(4)&7.6(-6)\tablenotemark{c} & 7 \\
$[$SIII$]$ 33 $\mu$m&3.0(-13)&4.1(3)&7.6(-6)\tablenotemark{c}  & 7 \\
$[$SIV$]$ 10.4 $\mu$m&9.5(-14)&4.0(5)&7.6(-6)\tablenotemark{c}  & 8\\

\enddata
\tablenotetext{a,b,c}{Abundances are from a. Savage \& Sembach 1996,
b. Grevesse \& Sauval 1998
and c. Asplund et al. 2005}
\tablerefs{
1. Pelan \& Berrington 1995
2. Galavis, Mendoza \& Zeippen 1995
3. Lennon \& Burke 1994
4. Blum \& Pradhan 1992
5. Griffin et al. 2001
6. Butler \& Zeippen, 1994 
7. Tayal \& Gupta 1999
8. Tayal 2000}
\end{deluxetable}

\begin{figure}
\plotone{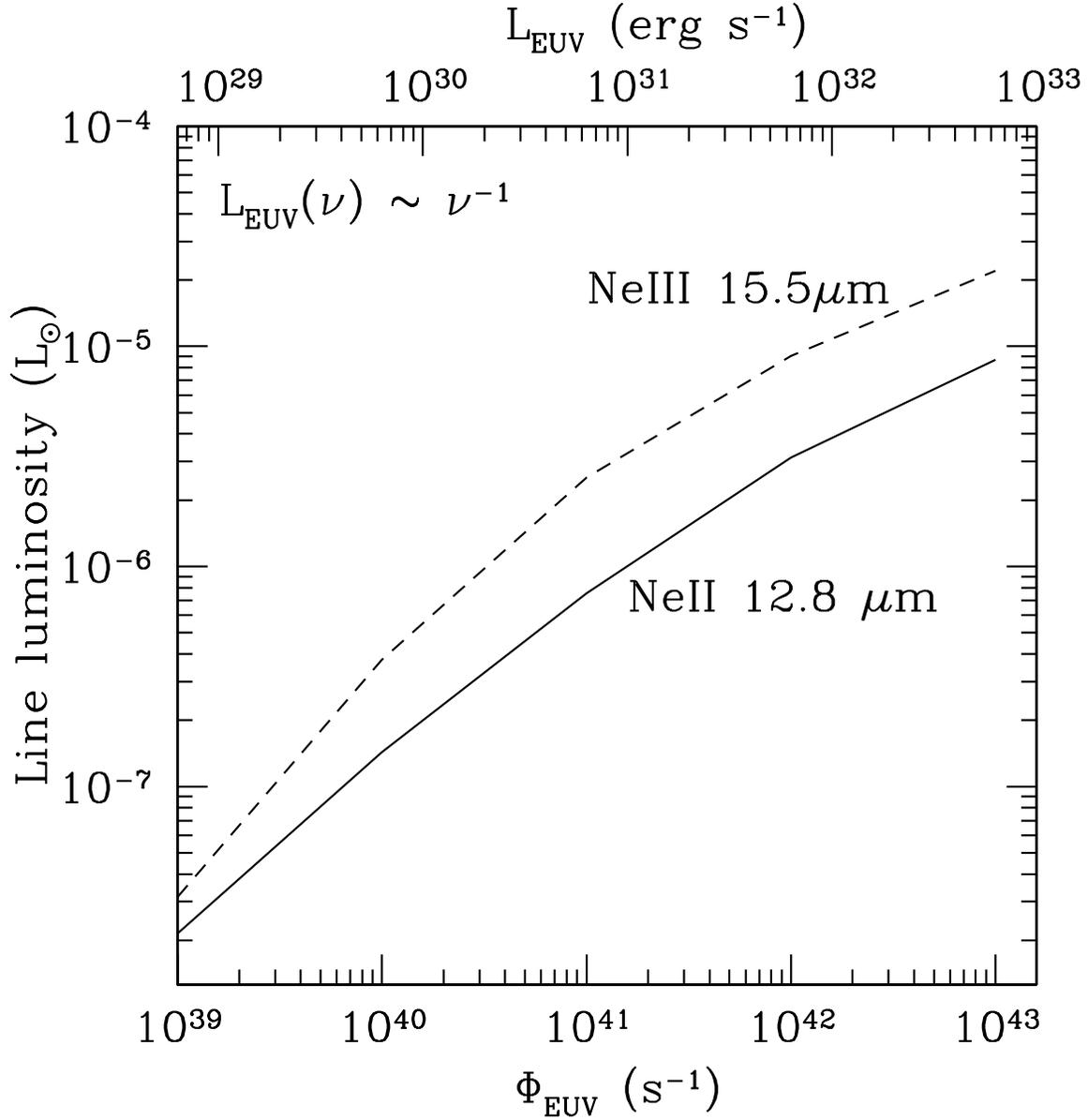}
\caption{The dependence of [NeII] 12.8 $\mu$m and [NeIII] 15.5 $\mu$m luminosity with the EUV
luminosity (top in erg s$^{-1}$ and bottom in EUV photons s$^{-1}$) of the central star.  The EUV spectrum
is assumed to be a power law, $L_{EUV}(\nu) \propto \nu ^{-1}$.   This relatively hard EUV spectrum
leads to high abundances of Ne$^{++}$ in the EUV layer, and [NeIII] stronger than [NeII].   In Section
3 we explain the overall dependence of line luminosity proportional to EUV luminosity with saturation
occurring at the higher luminosities as electron densities exceed the critical density of the [NeII] and 
[NeIII] transitions.}
\end{figure}

\begin{figure}
\plotone{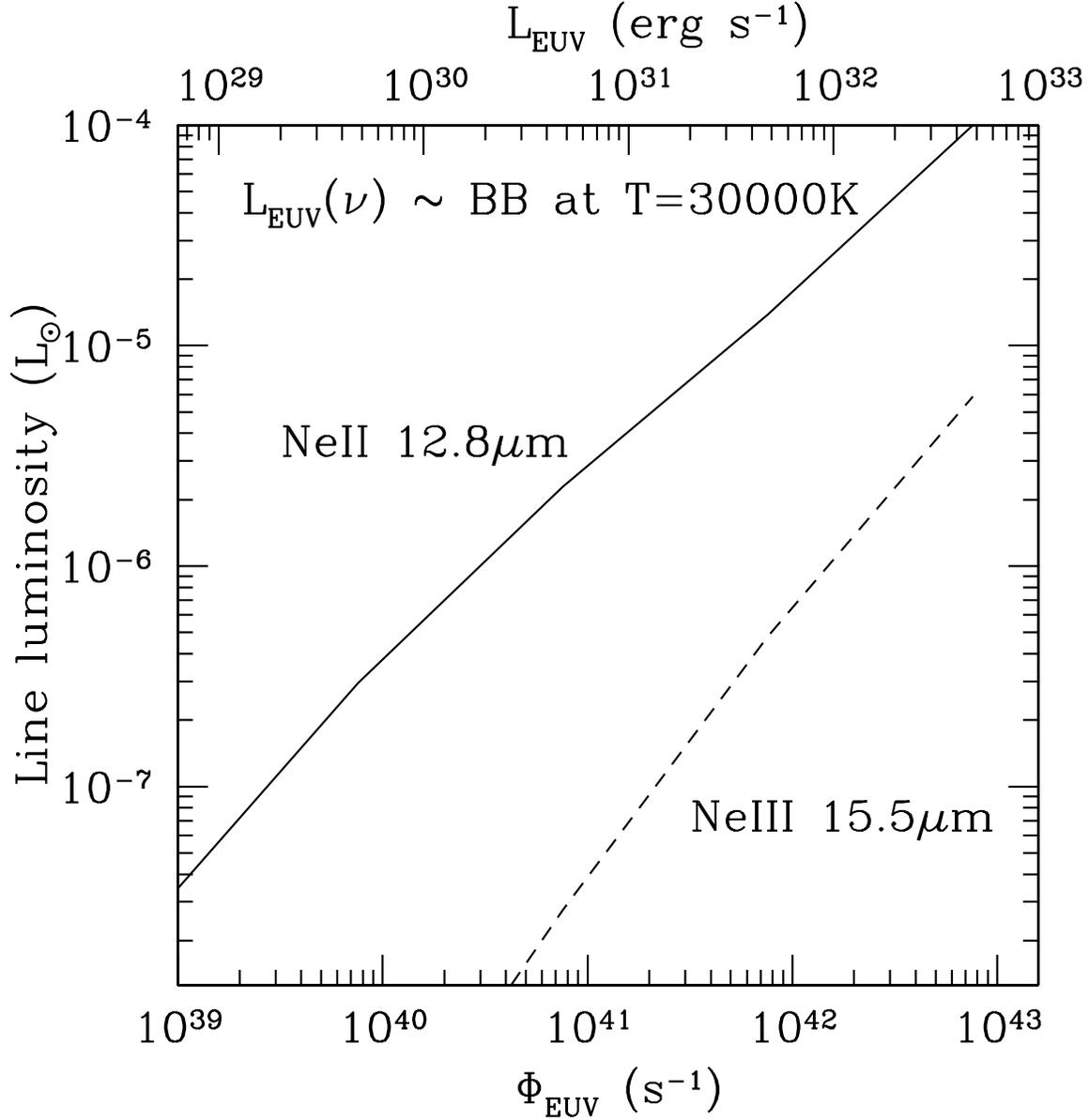}
\caption{The dependence of [NeII] 12.8 $\mu$m and [NeIII] 15.5 $\mu$m luminosity with the EUV
luminosity (top in erg s$^{-1}$ and bottom in EUV photons s$^{-1}$) of the central star.  The EUV spectrum
is assumed to be a blackbody with effective temperature $T_{eff} = 30,000$ K.   This relatively soft EUV spectrum
leads to high abundances of Ne$^{+}$ in the EUV layer, and [NeII] significantly stronger than [NeIII].   In Section
3 we explain the overall dependence of line luminosity proportional to EUV luminosity.}
\end{figure}

\begin{figure}
\plotone{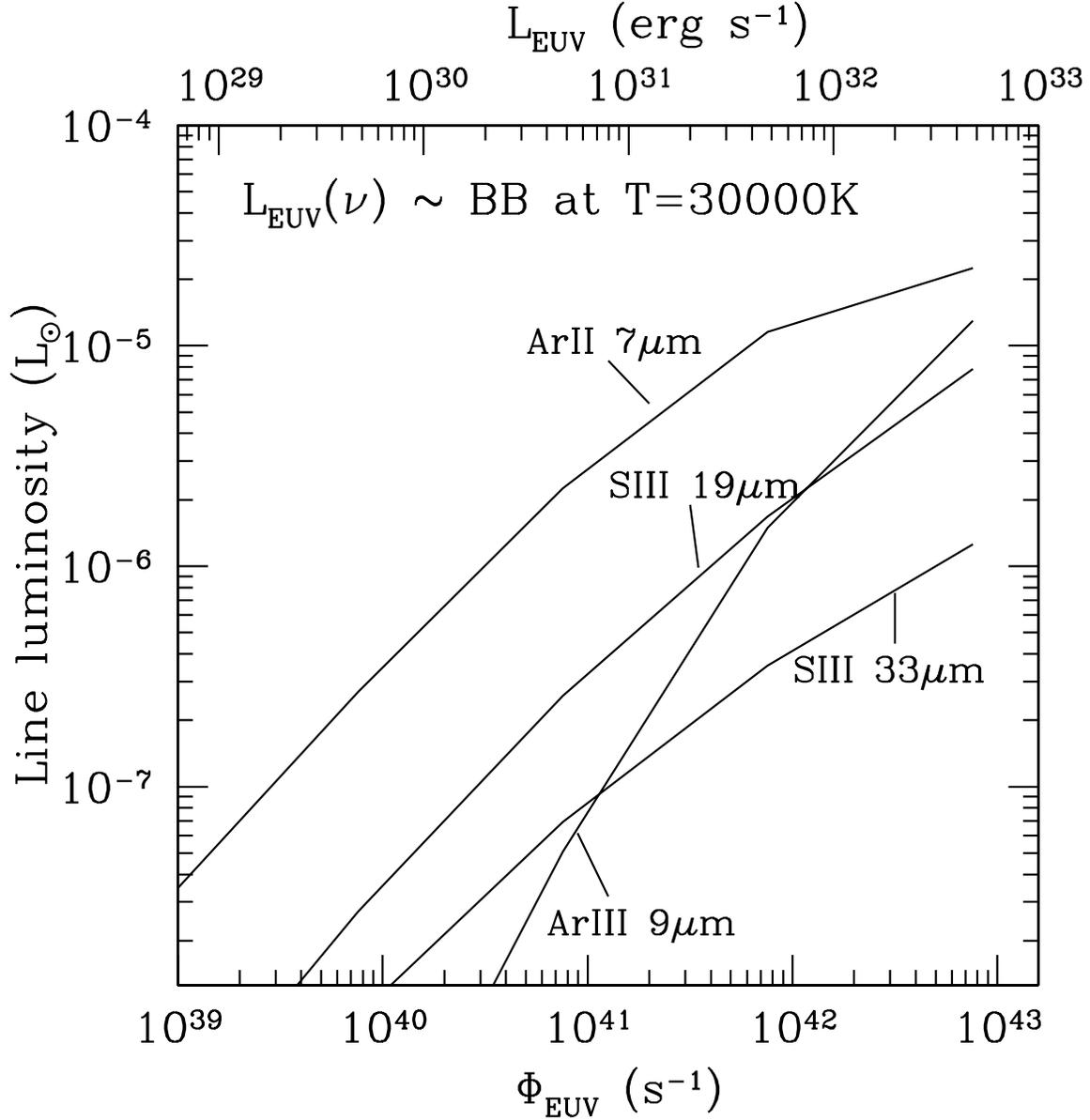}
\caption{The dependence of [ArII] 7 $\mu$m,  [ArIII] 9 $\mu$m,  [SIII] 19 $\mu$m, and
[SIII] 33 $\mu$m line luminosities on the EUV
luminosity (top in erg s$^{-1}$ and bottom in EUV photons s$^{-1}$) of the central star.  The EUV spectrum
is assumed to be a blackbody with effective temperature $T_{eff} = 30,000$ K.  Other lines from ionized species
which require $>13.6$ eV for their ionization are significantly weaker (see Table 1 and Eq. 8).   We discuss analytic 
approximations
for these predicted line luminosities in Section 3.1. }
\end{figure}

\begin{figure}
\plotone{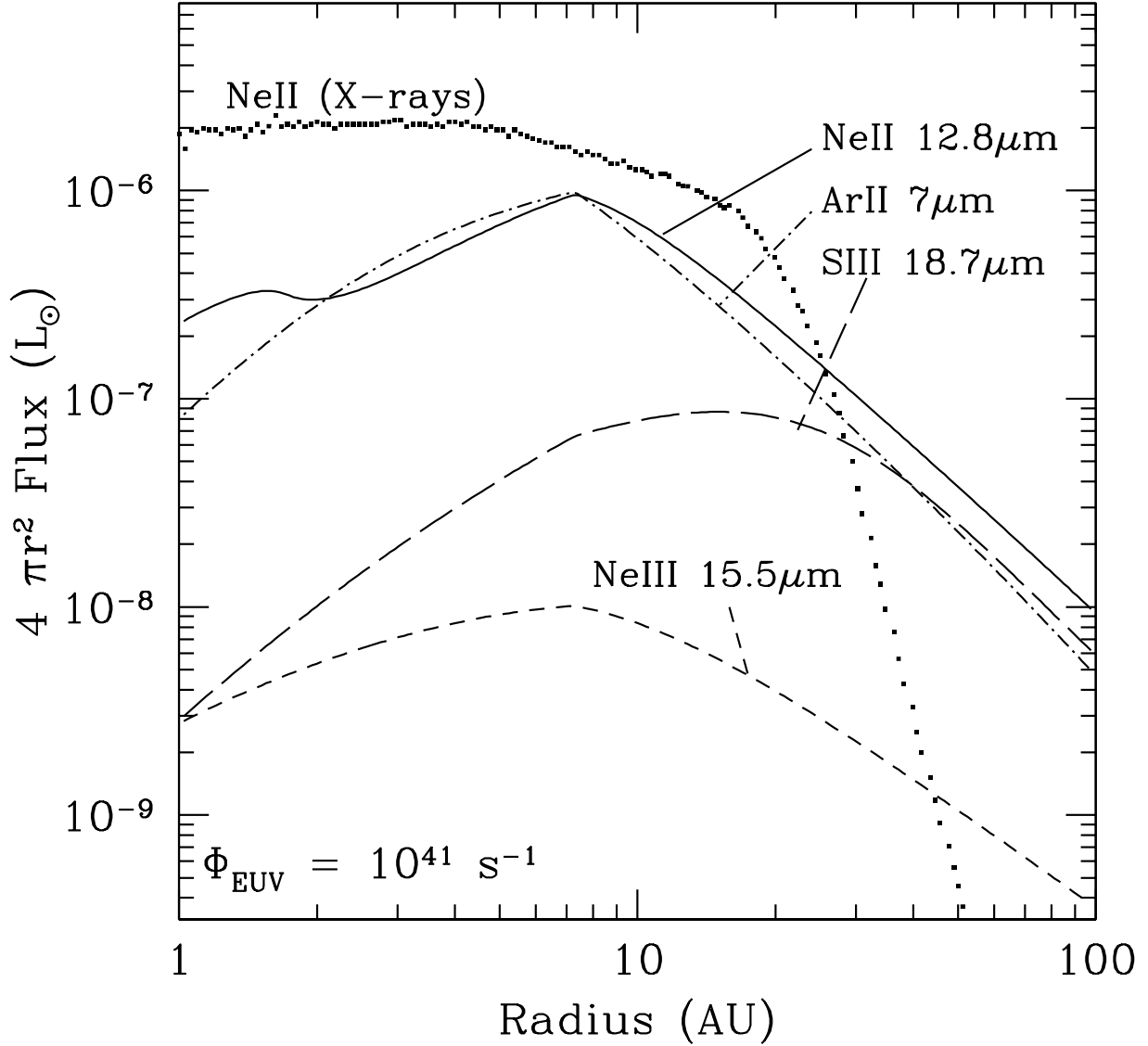}
\caption{The EUV-produced line flux emergent from one side of the disk times $4\pi r^2$ is plotted against the radius of
the disk.   This luminosity is approximately the luminosity emerging from both sides of  an annulus between $0.5r$
and $1.5r$.   The figure shows that most of the luminosity is generated at $r\sim 10$ AU.       Shown are the results for a central star
with $L_{EUV} \simeq  L_X = 2 \times 10^{30}$ erg s$^{-1}$.    In photon units, $\Phi _{EUV} = 10^{41}$ s$^{-1}$.
The EUV spectrum is assumed
to be a blackbody with effective temperature of 30,000 K (same case as Fig. 2). 
In addition, we have plotted (dotted line) $4\pi r^2$ times the emergent flux of [NeII] from the X-ray layer for
our standard ("hard") X-ray spectrum.  Substantial luminosity emerges from the region $\lta 1$ to 10 AU,
and the overall [NeII] luminosity is $\sim 2$ times greater than the EUV layer in this case.  }
\end{figure}

\begin{figure}
\plotone{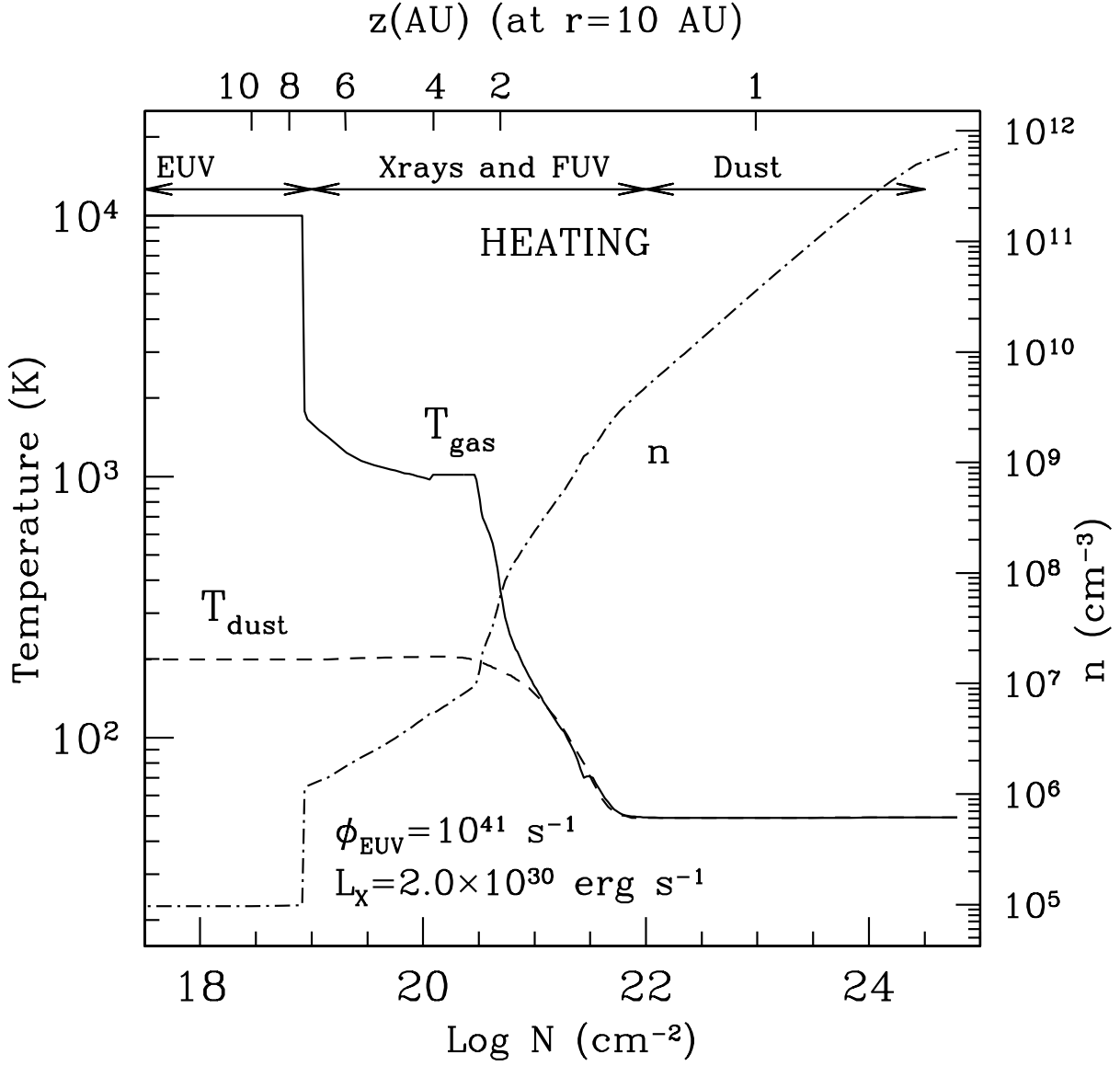}
\caption{ The gas temperature $T_{gas}$, the dust temperature $T_{dust}$, and the hydrogen nucleus
density $n$ are plotted versus the vertical distance $z$ (top) from the midplane, or the hydrogen 
nucleus column $N$ (bottom) from the surface ($z=r$).  This vertical slice is for $r= 10$ AU.  The central star X-ray luminosity and EUV luminosity
and spectrum are the same as in Figure 4. Note that $T_{gas}$ tracks $T_{dust}$ to $z\sim 2$ AU, or
$N\sim 10^{21}$ cm$^{-2}$.   Higher in the disk, the gas is hotter than the dust.  The EUV and X-ray layers
are marked.  Note that the ionization front is at $z_{IF} \simeq 7.5$ AU.  Dust dominates the heating of the gas near the midplane.}
\end{figure}

\begin{figure}
\plotone{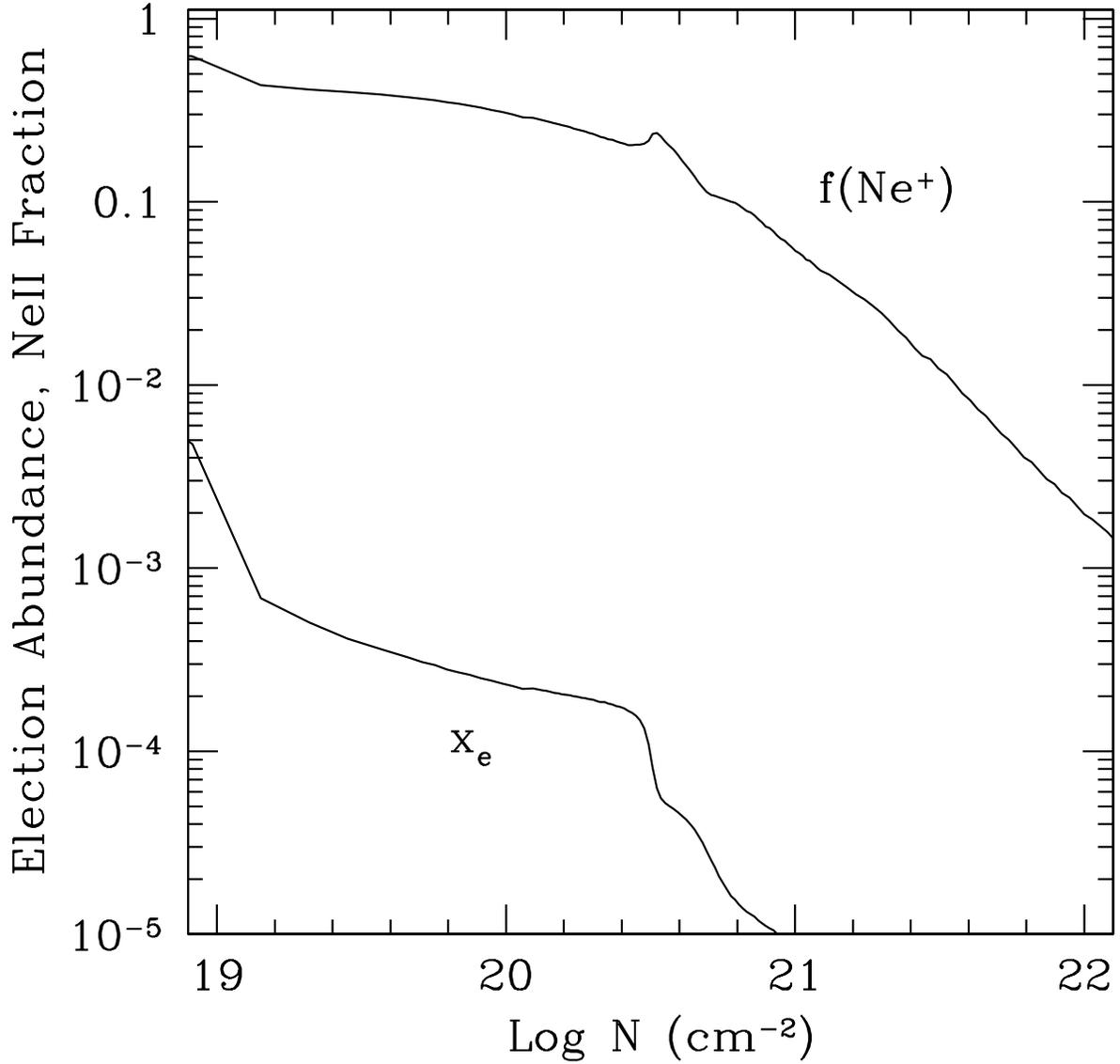}
\caption{The fraction of neon which is in the singly ionized state, $f(Ne^+)$, and the electron abundance
relative to hydrogen nuclei, $x_e$, are plotted versus the hydrogen nucleus column $N$ from the disk
surface at $r= 10$ AU.   The EUV layer only extends to 
about $N \sim 10^{19}$ cm$^{-2}$ (see Figure 5) so that we highlight
here the X-ray layer.  The X-ray spectrum is our harder spectrum which peaks at $\sim 2$ keV.  X-rays 
maintain the high $f(Ne^+)$ throughout the region plotted.   X-rays maintain
a relatively high electron abundance to $N\sim 10^{20}$ cm$^{-2}$.  At higher columns, FUV photoionization
of carbon as well as X-rays maintain $x_e$. }
\end{figure}

\begin{figure}
\plotone{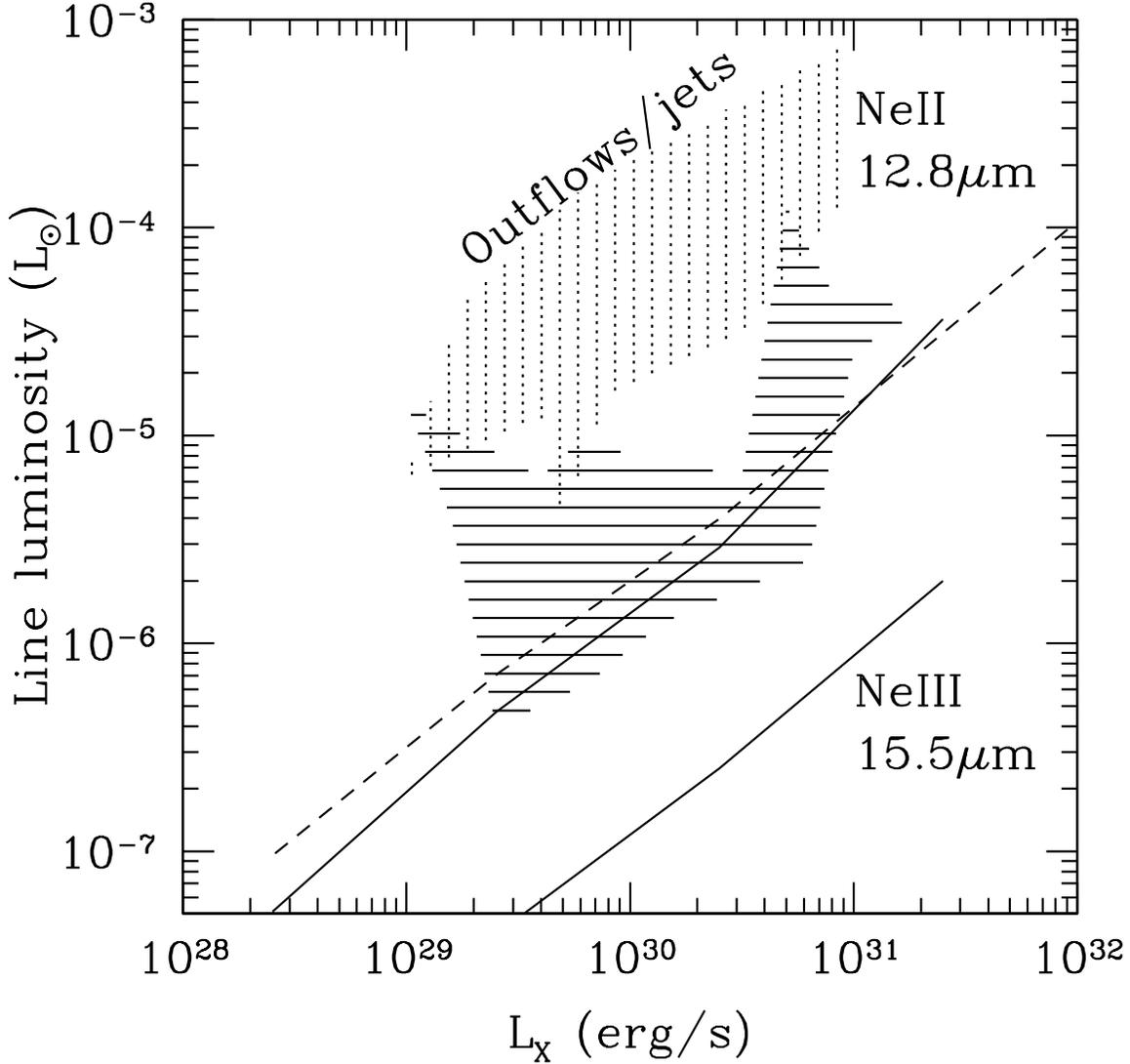}
\caption{The dependence of [NeII] 12.8 $\mu$m and [NeIII] 15.5 $\mu$m luminosity with the X-ray 
luminosity.  The solid labelled lines are [NeII] and [NeIII] for our "harder" X-ray spectrum, where $L_\nu
\propto \nu$ for 0.1 keV$< h\nu < 2$ keV.   The dashed line is the [NeII] luminosity for our softer X-ray spectrum
source, where $L_\nu \propto \nu^{-1}$ for 0.1 keV $< h\nu < 2$ keV. Note the nearly linear dependence $L_{[NeII]}$
with $L_X$.  Comparison with Fig. 2 shows that if $L_X \sim L_{EUV}$, and assuming a soft EUV spectrum that produces the maximum amount of [NeII], then the [NeII] line luminosity is still 2 times stronger from
the X-ray layer as from the EUV layer.  Also plotted are a recent compilation of [NeII] and $L_X$ data (G\"udel et al 2009).  The region shaded with vertical dotted lines are sources
with known outflows or jets.  The region with horizontal solid lines are sources with undetected outflows.
 It appears that there are a substantial number of sources, especially the "outflow/jet"
sources that are more luminous in [NeII] than the X-ray layer (or the EUV layer) could provide; internal shocks
in the winds or jets are a possible explanation for these sources.  }
\end{figure}

\begin{figure}
\plotone{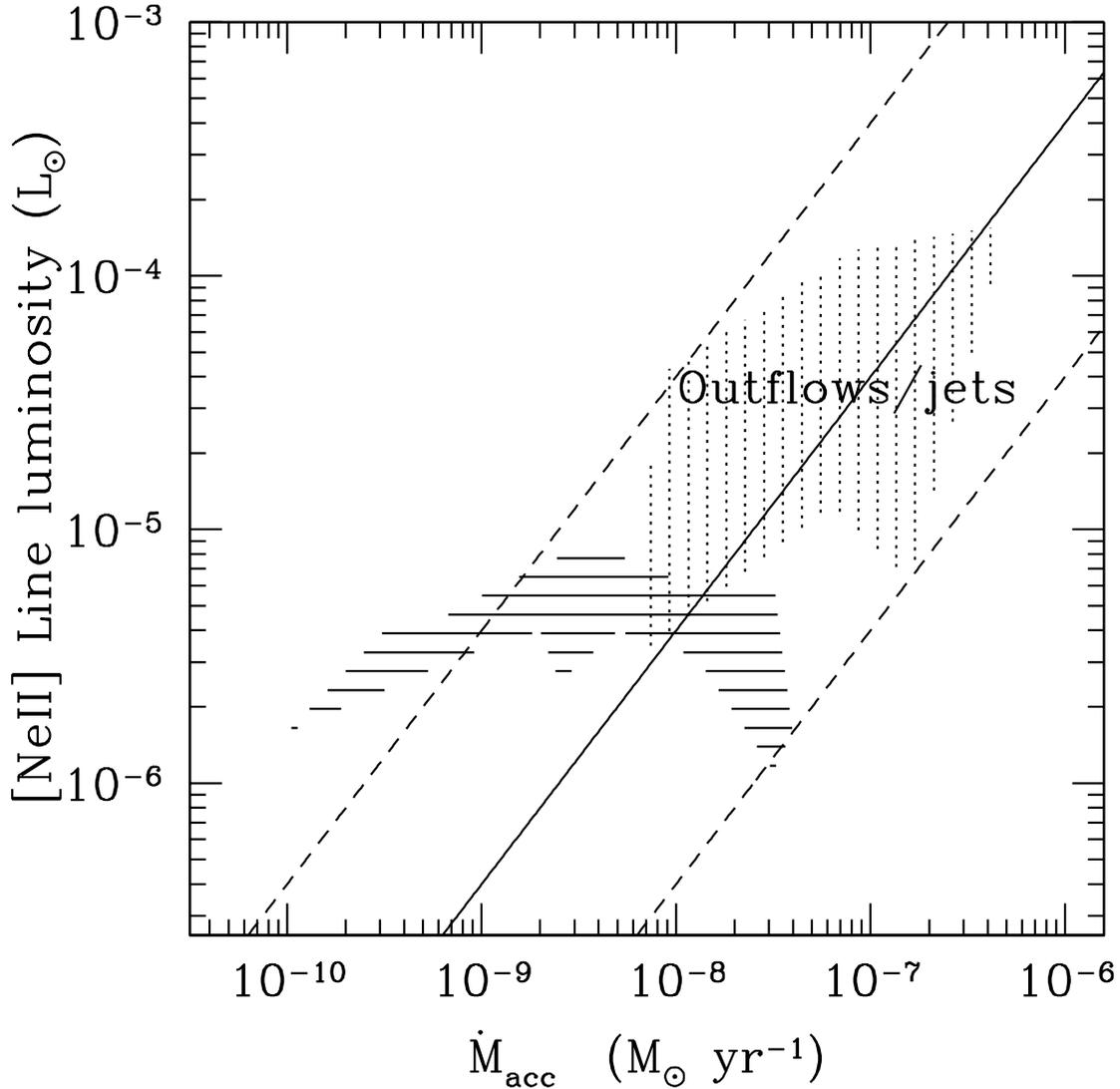}
\caption{The dependence of [NeII] 12.8 $\mu$m luminosity with the mass accretion rate onto the
central star. The shaded regions are the same notation as in Figure 7; data from G\"udel et al (2009).   Section 3.4 in the text and Eq. (33) predict the [NeII] luminosity as a function of the wind or jet mass loss rate $\dot M_w$ and
the fraction $f_{sh}$ of the wind or jet that shocks at speeds greater than about 100 km s$^{-1}$. 
The solid line in the figure assumes $\dot M_w = 0.1 \dot M_{acc}$ and $f_{sh}= 1$.  The upper dashed  line assumes
 $\dot M_w =  \dot M_{acc}$ and $f_{sh}= 1$.  The lower dashed line assumes that the product
 $f_{sh}\dot M_w$ is 10 times less than assumed in the solid line case.   Shocks appear viable explanations
 for the origin of many of the [NeII] sources, especially those with observed outflows and jets (see also,
 Figure 7). }
\end{figure}

\begin{figure}
\plotone{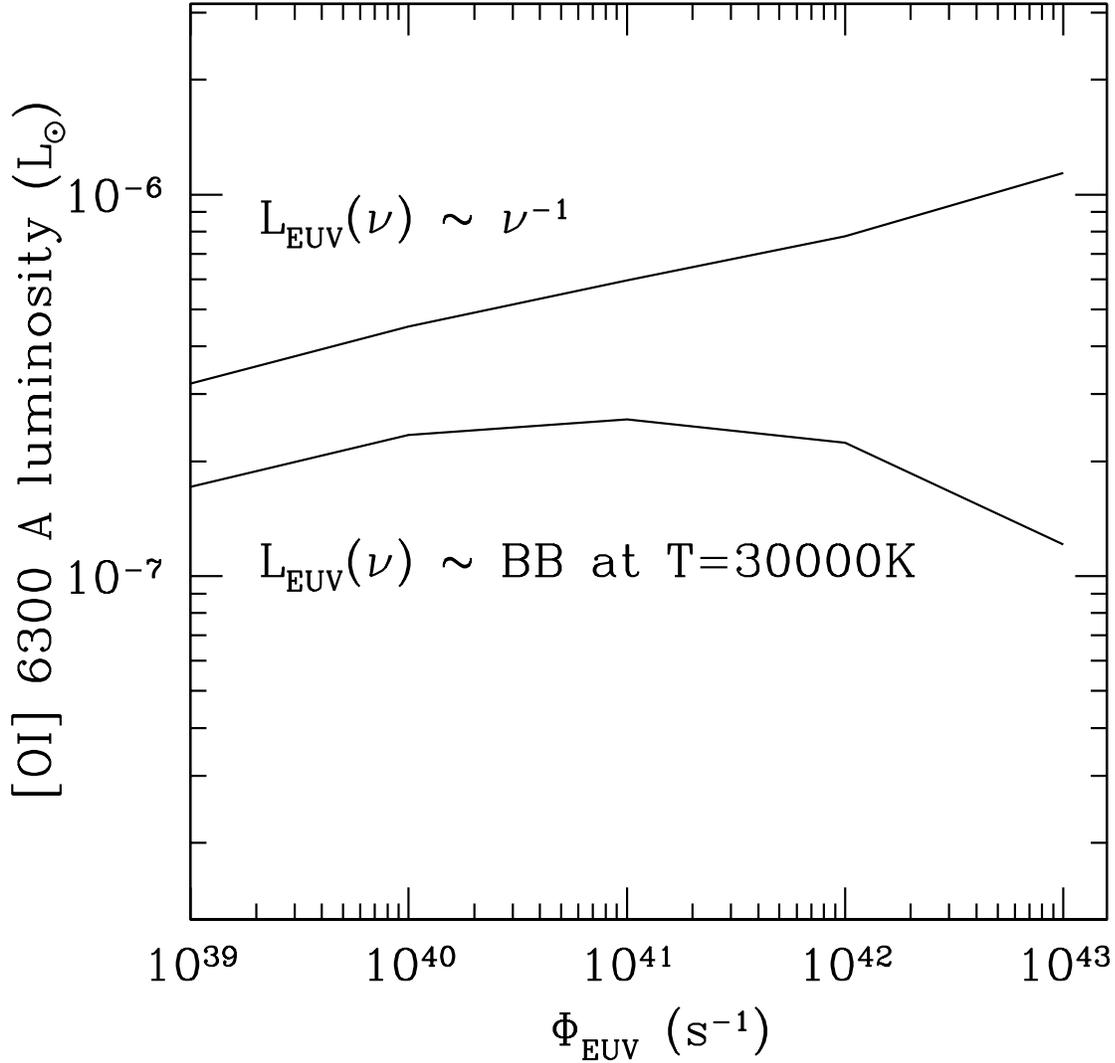}
\caption{The predicted [OI] 6300 \AA \ luminosity from the EUV layer is plotted for both a blackbody
EUV spectrum ($T_{eff} = 30,000$ K) or for a power law spectrum as a function of the EUV photon
 luminosity $\Phi _{EUV}$.  The harder spectrum produces more [OI] luminosity because more
 atomic O survives in the mostly ionized EUV layer (see text). Observed [OI] luminosities are typically
 much higher than the $\lta 10^{-6}$ L$_\odot$ predicted from the EUV layer (see text).}
\end{figure}

\begin{figure}
\plotone{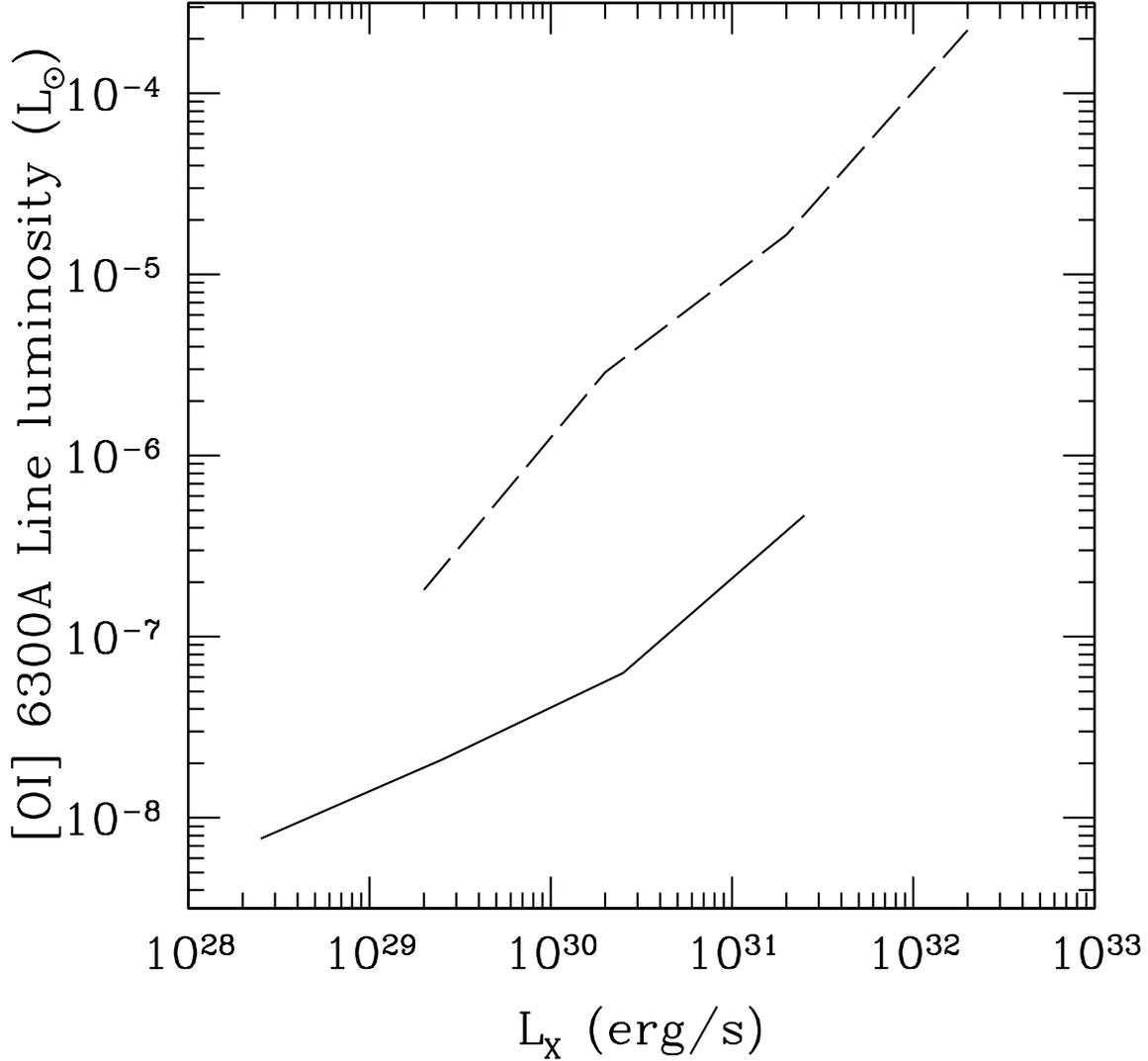}
\caption{The predicted [OI] 6300 \AA \ luminosity from the X-ray layer is plotted versus the X-ray 
luminosity of the central star.  The solid line is for our harder X-ray spectrum whereas the dashed line
is for our softer X-ray spectrum (see text or caption to Fig. 7).  Observed [OI] luminosities are typically
 much higher than the $\lta 10^{-6}$ L$_\odot$ predicted from the X-ray layer produced by the harder
 spectrum.   However, the softer X-ray spectrum produces [OI] luminosities much more in accord
 with observations, because the X-ray layer is warmer and the line is extremely temperature
 sensitive (see text).  }
\end{figure}

\end{document}